\documentclass[useAMS,usenatbib]{mn2e}

\usepackage{graphicx}
\usepackage{layouts}
\usepackage{pdflscape}

\def \Ha {H$\alpha$\ }

\def \apjs {ApJS}
\def \aj {AJ}
\def \apj {ApJ}
\def \pasp {PASP}
\def \araa {ARA\&A}
\def \mnras {MNRAS}
\def \aap {A\&A}
\def \apjl {ApJ}
\def \nat {Nature}
\def \aaps {A\&AS}
\def \physrep{Physics Reports}

\def\lesssim{\mathrel{\hbox{\rlap{\hbox{\lower4pt\hbox{$\sim$}}}\hbox{$<$}}}}
\def\gtrsim{\mathrel{\hbox{\rlap{\hbox{\lower4pt\hbox{$\sim$}}}\hbox{$>$}}}}

\newcommand\ion[2]{#1$\,${\scshape{#2}}}

\makeatletter
\def\@xfootnote[#1]{%
 \protected@xdef\@thefnmark{#1}%
 \@footnotemark\@footnotetext}
\makeatother

\title[The multi-faceted SN 2014G]{The multi-faceted Type II-L supernova 2014G from pre-maximum to nebular phase}
\author[G. Terreran]{G. Terreran$^{1,2}$\thanks{E-mail:gterreran01@qub.ac.uk},
A. Jerkstrand$^{1}$, S. Benetti$^{2}$, S.~J. Smartt$^{1}$, P. Ochner$^{2}$, 
\newauthor L. Tomasella$^{2}$, D.~A. Howell$^{3,4}$, A. Morales-Garoffolo$^{5}$, A. Harutyunyan$^{6}$, 
\newauthor E. Kankare$^{1}$, I. Arcavi$^{3,7}$, E. Cappellaro$^{2}$, N. Elias-Rosa$^{2}$, G. Hosseinzadeh$^{3,4}$, 
\newauthor T. Kangas$^{8}$, A. Pastorello$^{2}$, L. Tartaglia$^{2,9}$, M. Turatto$^{2}$, S. Valenti$^{10}$, 
\newauthor P. Wiggins$^{11}$, F. Yuan$^{12,13}$.\\
$^{1}$Astrophysics Research Centre, School of Mathematics and Physics, Queen's University Belfast, Belfast BT7 1NN, UK\\
$^{2}$INAF-Osservatorio Astronomico di Padova, Vicolo dell'Osservatorio 5, I-35122 Padova, Italy\\
$^{3}$Las Cumbres Observatory, Global Telescope Network, 6740 Cortona Drive Suite 102, Goleta, CA 93117, USA\\
$^{4}$Department of Physics, University of California, Santa Barbara, Broida Hall, Mail Code 9530, Santa Barbara, CA 93106-9530, USA\\
$^{5}$Institut de Ci\`encies de l'Espai (CSIC-IEEC), Campus UAB, Cam\`i de Can Magrans S/N, 08193 Cerdanyola (Barcelona), Spain\\
$^{6}$Telescopio Nazionale Galileo, Fundaci\'on Galileo Galilei - INAF, Rambla Jos\'e Ana Fern\'andez P\'erez, 7, E-38712 Bre\~na Baja, TF, Spain\\
$^{7}$Kavli Institute for Theoretical Physics, University of California, Santa Barbara, CA 93106, USA\\
$^{8}$Tuorla Observatory, Department of Physics and Astronomy, University of Turku, V\"ais\"al\"antie 20, 21500 Piikki\"o, Finland\\
$^{9}$Universit\`a degli Studi di Padova, Dipartimento di Fisica e Astronomia, Vicolo dell'Osservatorio 2, 35122 Padova, Italy\\
$^{10}$Department of Physics, University of California, Davis, CA 95616, USA\\
$^{11}$Wiggins Observatory, Tooele, UT 84074, USA\\
$^{12}$Research School of Astronomy and Astrophysics, Australian National University, Canberra, ACT 2611, Australia 2ARC\\
$^{13}$Centre of Excellence for All-sky Astrophysics (CAASTRO)}

\begin{document}


\pagerange{\pageref{firstpage}--\pageref{lastpage}} \pubyear{2014}

\maketitle

\label{firstpage}

\begin{abstract}
We present multi-band ultraviolet, optical, and near-infrared photometry, along with visual-wavelength spectroscopy, of supernova (SN) 2014G in the nearby galaxy \mbox{NGC 3448} (25\,Mpc). The early-phase spectra show strong emission lines of the high ionisation species He II/N IV/C IV during the first 2-3~d after explosion, traces of a metal-rich CSM probably due to pre-explosion mass loss events. These disappear by day 9 and the spectral evolution then continues matching that of normal Type II SNe. The post-maximum light curve declines at a rate typical of \mbox{Type II-L} class. The extensive photometric coverage tracks the drop from the photospheric stage and constrains the radioactive tail, with a steeper decline rate than that expected from the $^{56}$Co decay if $\gamma$-rays are fully trapped by the ejecta. We report the appearance of an unusual feature on the blue-side of H$\alpha$ after 100~d, which evolves to appear as a flat spectral feature linking \Ha and the [\ion{O}{I}] doublet. This may be due to interaction of the ejecta with a strongly asymmetric, and possibly bipolar CSM. Finally, we report two deep spectra at $\sim190$ and 340~d after explosion, the latter being arguably one of the latest spectra for a Type II-L SN. By modelling the spectral region around the [\ion{Ca}{II}], we find a supersolar Ni/Fe production. The strength of the [\ion{O}{I}] $\lambda\lambda$6300,6363 doublet, compared with synthetic nebular spectra, suggests a progenitor with a zero-age main-sequence mass between 15 and 19 M$_\odot$.
\end{abstract}


\section{Introduction}\label{sec: introduction}
The classification of supernovae (SNe) is mainly based on observational features. The presence of hydrogen in the spectrum primarily splits SNe into Type I (hydrogen-poor) and Type II (hydrogen-rich). 
Focusing on Type II SNe, the amount of hydrogen that the progenitor stars retained at the time of the explosion strongly affects the evolution of the SN and in particular the shape of the light curve (LC). It can show a plateau (Type II-P SNe) or a linear decline after peak (Type II-L SNe). Type II-P SNe are supposed to arise from progenitor stars in the mass range of $8-17$ M$_\odot$ \citep{Smartt 2009} that reach the evolution state at which the collapse of the iron core occurs. It has also been suggested that the progenitors of Type II-L SNe could have larger masses at zero-age main-sequence (ZAMS) than Type II-P \citep{Elias-Rosa 2010,Elias-Rosa 2011,Anderson 2012,Kuncarayakti 2013}. In particular, Type II-L SNe are thought to originate from progenitor stars with less hydrogen ($1-2$ M$_\odot$) and larger radii (few 1000 R$_\odot$) with respect to those that give birth to Type II-P SNe \citep{Swartz 1991,Blinnikov 1993,Popov 1993,Arnett 1996}. It has been proposed a physical continuity from Type II-L to Type IIb SNe, which show even less amount of hydrogen in the spectra at early phases, and then to Type I-b SNe, where there is no sign of hydrogen \citep{Nomoto 1993}.

Whether Type II-P and II-L SNe are two clearly separated subclasses, and thus originated from two different types of progenitors, or a continuum exists between the two is still debated \citep{Patat 1994,Arcavi 2012,Anderson 2014a,Faran 2014a,Faran 2014b,Gall 2015,Gaitan 2015,Poznanski 2015,Sanders 2015,Galbany 2016,Valenti 2016}. The lack of significant spectroscopic differences between the two classes leaves just the LC shape as the discriminating factor (or more quantitatively, the luminosity decline rate). In the literature, there are several parameters proposed for a quantitative discrimination between the Type II-P and II-L SNe, based on different phase intervals for measuring the average decline rate in different optical bands, e.g., 3.5 mag~100~d$^{-1}$ in $B$-band ($B_{100}$) by \cite{Patat 1994}, 0.5 mag~50~d$^{-1}$ in $V$-band ($V_{50}$) by \cite{Faran 2014b}, again 0.5 mag~50~d$^{-1}$ but in $R$-band ($R_{50}$) by \cite{Li 2011}.

In addition to these sub-classes, Type IIn SNe are distinguished due to their evolution being significantly affected by the presence of circumstellar material (CSM) in the proximity of the star. This CSM gets shocked by the ejecta, producing narrow emission lines superimposed on the SN spectrum. These narrow lines are the traits that typically identify this subclass. Thus the Type IIn SNe are classified as such purely based on their spectroscopic features. 
The LCs of Type IIn SNe are usually quite heterogeneous, and depending on the masses, density and geometry of the CSM, they can fall linearly or they can stay bright even for many years \citep[see][]{Taddia 2015}. The observed CSM can be generated by pre-SN mass loss events, typical of massive progenitors \citep{Smartt 2009} or binary systems \citep{Chevalier 2012}, or it can also be the result of almost static, photoionisation-confined shells created by the stellar wind of lower-mass red supergiants \citep[RSGs,][]{Mackey 2014}. Thus the progenitors of Type IIn SNe could conceivably be spread over a wide range of masses and environments. For a more thorough description on the SNe classes and subclasses, including the hydrogen-poor events, see \cite{Filippenko 1997} and \cite{Turatto 2007}.

There are only few Type II-L SNe which are nearby (less than about 30 Mpc), have been discovered shortly after the explosion, and were monitored until the nebular phase. SN 2014G is a new entry that will enrich the statistics for the Type II-L SNe. At coordinates \mbox{$\alpha\,=\,10^{\rm{h}}54^{\rm{m}}34^{\rm{s}}.1$}, \mbox{$\delta\,=\,+54\degr17\arcmin56\farcs9$}, it was discovered in the nearby galaxy NGC 3448 independently by two amateur astronomers, P. Wiggins and K. Itagaki, with the first detection on 2014 January 14.32 UT \citep{Nakano 2014}, and was initially reported in CBAT Transient Object Followup Reports\footnote{http://www.cbat.eps.harvard.edu/unconf/tocp.html} as PSN J10543413+5417569. A useful constraint on the explosion epoch was given by the Master collaboration \citep{Lipunov 2010} which reported an unfiltered limit magnitude of 19.4 in a combination of 6 exposures of the host galaxy taken on January $10.85-10.88$ UT, during routine survey mode. We then set the explosion epoch on January 12.6 UT (MJD 56669.6), with an uncertainty of $\pm$1.7~d. The epoch of explosion will be taken as reference throughout the paper. Our group took a classification spectrum less than a day after discovery \citep{Ochner 2014}. This showed several narrow features superimposed on a blue continuum and a comparison made with GELATO \citep{Harutyunyan 2008} gave the best match with a Type IIn SN. In the following days these narrow lines disappeared and the spectra started to resemble a more normal Type II SN. The photometric evolution, on the other hand, showed a monotonic decline, which led to revise the classification to a Type II-L \citep{Eenmae 2014}.

At the time of writing, another paper on SN 2014G was published \citep[][hereafter B16]{Bose 2016}. They presented the photometry and polarimetry of the transient, but no spectroscopy. Their work will be used as a useful comparison with our data and conclusions. Although of minimum relevance, we point out that they set the explosion epoch on 2014 January 12.2 UT, so a minor 0.4~d discrepancy arises between our and their phase reference.

In the following, we will first briefly describe the host galaxy in Section \ref{NGC}. In Section \ref{Observations} we describe the instrumentation and the reduction techniques used. We will then present the photometric and the spectroscopic evolution in Sections \ref{Photometry} and \ref{Spectroscopy} respectively. Finally we will discuss the results in Section \ref{Discussion}, focusing on the physical interpretation of the data for this transient.

\section[The host galaxy]{The host galaxy}\label{NGC}

\begin{figure*}
\includegraphics[width=168mm]{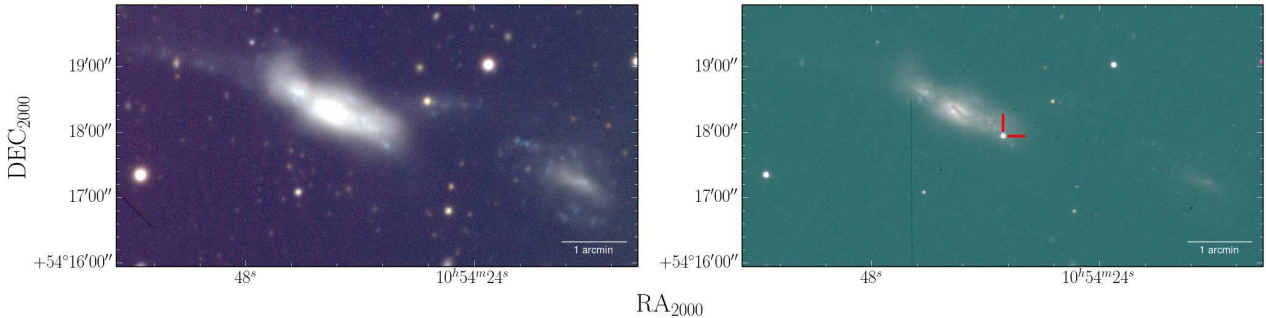}
\caption{RGB image of the SN 2014G field. On the left, a reference image where the whole system Arp 205 (NGC 3448+UGC 6016) is clearly visible. On the right, a post explosion image, where SN 2014G is marked in red.}
\label{fig: rgb}
\end{figure*}

NGC 3448 is the host galaxy of SN 2014G and together with the dwarf companion UGC 6016 (visible in the left panel of Figure \ref{fig: rgb}) they make the system Arp 205 \citep{Arp 1966}. The morphology of NGC 3448 shows signatures of a tidal interaction with its companion \citep{Noreau 1986}. It belongs to the amorphous class of galaxies \citep{Sandage 1979}, characterised by a smooth appearance and by their high star-formation rates. The main structure is an edge-on disc, with two bright bulges and an optically-thick dust line crossing the central one (see right panel of Figure \ref{fig: rgb}). \cite{Bertola 1984} studied the optical rotation curve of NGC 3448 showing heliocentric velocities ranging from 1150\,km~s$^{-1}$ in the SW region up to 1400\,km~s$^{-1}$ in the NE region. Two higher-velocity clumps departing from the ``rigid'' component were also identified and interpreted as foreground gas clouds of tidal origins and now infalling towards the galaxy. Several knots and unresolved radio sources can also be seen which are probably associated with a complex of SN remnants \citep{Noreau 1987}. Ultraviolet (UV) and radio analyses showed that NGC 3448 is indeed a starburst galaxy. By a comparison with the nearby starburst galaxy M82, a SN rate of one event every $\sim10$ years was estimated. More recent works produced drastically lower
star formation rate of 1.4 M$_\odot$~yr$^{-1}$ \citep{Lanz 2013}, which translates to just 0.01 SN per year.

SN 2014G is the first SN reported in NGC 3448. It resides 44\arcsec\ west and 20\arcsec\ south of the centre \citep{Nakano 2014} and then, correcting for the rotation curve in \cite{Bertola 1984}, it should have a heliocentric velocity of \hbox{1175\,km~s$^{-1}$}. 
This value is in good agreement with the value we measured from the narrow lines visible in the early spectra. However as the host is edge-on, and being the SN in the external part of the galaxy, its proper motion could lead to a miscalculation of the distance. Thus, we prefer to use a redshift-independent measure \citep[using the Tully-Fisher method,][]{Tully 1988} of the distance of NGC 3448 reported in the NASA Extragalactic Database\footnote{http://ned.ipac.caltech.edu} (NED), which is $\umu\simeq31.94\pm0.80$ \citep{Tully 1988}, equivalent to a distance of $\sim24.5$ Mpc. This value will be used through the paper.

The Galaxy reddening along the line of sight from the all-sky Galactic dust-extinction survey is \hbox{E($B-V$)$_\rmn{Gal}$=0.01}~mag \citep{Schlafly 2011}. A narrow \hbox{Na I D} absorption line is clearly visible in the first ten spectra and this can be used to estimate the reddening due to the host. Averaging the equivalent width (EW) of the doublet, we measured EW$\simeq1.33$~\AA. Following the lower relation in \cite{Turatto 2003} it yields \mbox{E($B-V$)$_\rmn{host}=0.20\pm0.11$~mag}. Thus a total reddening for SN 2014G \mbox{E($B-V$)$_\rmn{tot}=0.21\pm0.11$~mag} was assumed. B16 implemented the colour method \citep{Olivares 2010} to estimate the local reddening and came up with a total reddening of E($B-V$)$_\rmn{tot}=0.25$~mag, which is in good agreement with the one inferred here. 

\begin{table*}
 \centering
 \begin{minipage}{140mm}
 \caption{Instrumental configurations used for the follow-up campaign of SN 2014G}
 \begin{tabular}{@{}llcr@{$\times$}lcr@{ $-$ }l@{}}
 \hline
 Telescope & Location & Instrument & \multicolumn{2}{c}{Field of View} & Imaging & \multicolumn{2}{c}{Spectroscopy}\\
   &  &  & \multicolumn{2}{c}{ } & bands\footnote{Uppercase indicates Johnson-Cousins filters. Lowercase indicates SDSS filters.} & \multicolumn{2}{c}{wavelength range}\\
 \hline
Celestron C14 reflector\footnote{Amateur equipment operated by P. Wiggins}&near Erda, UT, USA&SBIG\footnote{SBIG ST-10XME}&18\arcmin& 26\arcmin & unfiltered\\
 Schmidt 67/92cm& Stazione osservativa di & SBIG\footnote{SBIG STL-11000MC2, donation from Rotary Club Asiago for outreach} & 58\arcmin&38\arcmin&\textit{BVRI}\\
 &Asiago Cima Ekar\\
 1.22m Galileo & Osservatorio astrofisico & Andor\footnote{Andor iDus DU440} & \multicolumn{2}{c}{ }&&3800&8000~\AA\\
 &di Asiago\\ 
  1.82m Copernico & Stazione osservativa di & AFOSC& 8\farcm7&8\farcm7&\textit{UBVRIgri}&3400&10000~\AA\\
 &Asiago Cima Ekar\\
 Liverpool Telescope& Observatorio del Roque & IO:O & 10\arcmin &10\arcmin & \textit{BVgriz}\footnote{Pan-STARRS \textit{z}} \\
 (LT) & de Los Muchachos\\
 Nordic Optical & Observatorio del Roque & StanCam& 3\arcmin&3\arcmin&\textit{BVRI}\\
 Telescope (NOT)&de Los Muchachos&NOTCam&4\arcmin&4\arcmin&\textit{JHK}\\
 &&ALFOSC&\multicolumn{2}{c}{ }&\textit{VRI}&3200&9100~\AA\\
 Telescopio Nazionale & Observatorio del Roque & DOLORES&8\farcm6&8\farcm6&\textit{UBVRIgri}&3000&10000~\AA\\
 Galileo (TNG)&de Los Muchachos\\
 Gran Telescopio & Observatorio del Roque&OSIRIS&8\farcm6&8\farcm6&&3600&7200~\AA\\
 CANARIAS (GTC) &de Los Muchachos\\
 LCOGT 1m & McDonald Observatory & SBIG\footnote{SBIG STX-16803} & 16\arcmin&16\arcmin& \textit{BVgri} \\
 LCOGT 2m & Haleakala Observatory & Spectral & 10\farcm5&10\farcm5 & \textit{gri} \\
 Telescopi Joan & Observatori Astron\`omic & MEIA & 12\farcm3 & 12\farcm3 & \textit{UBVRI} \\
 Or\'o (TJO) &del Montsec\\
 \textit{Swift} & & UVOT& 17\arcmin& 17\arcmin & \textit{UV-W1,M2,W2}\\
&&& \multicolumn{2}{c}{ }&\textit{UBV}\\
\hline
\end{tabular}
\label{tab: instr}
\end{minipage}
\end{table*}

\section[Observations]{Observations}\label{Observations}
\begin{figure*}
\includegraphics[width=168mm]{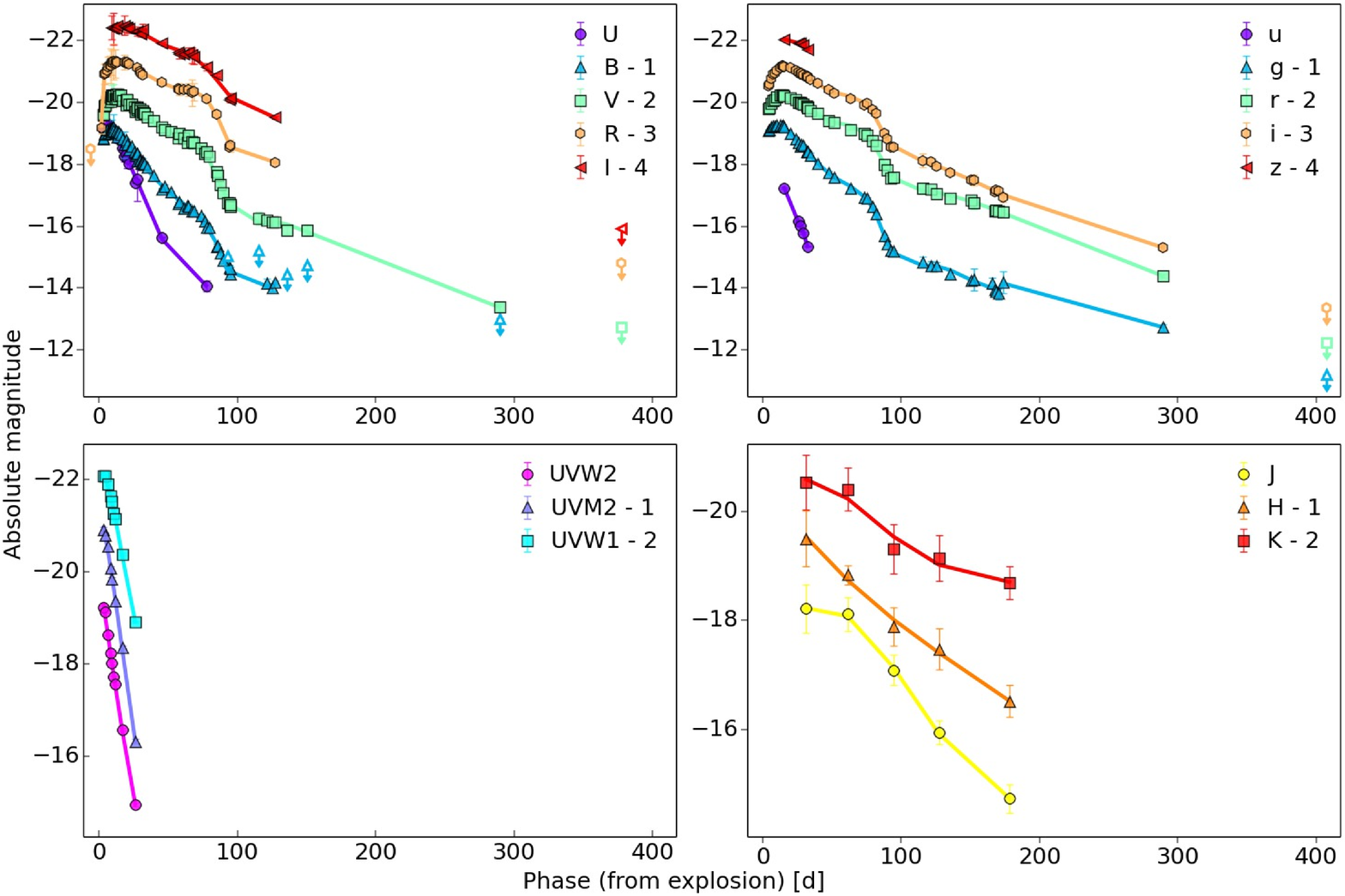}
\caption{Photometric evolution of SN 2014G. The optical bands in both Johnson/Cousins and SDSS filters are reported in top left and top right panels respectively. Note that the SDSS filters are calibrated in AB mag, while Johnson/Cousins in Vega mag. UV bands are shown in the bottom left panel and NIR in the bottom right. The unfiltered observations of P. W. were calibrated as Cousins $R$ band. Upper limits are indicated by an empty symbol with an arrow. The lines connecting the points are simple interpolations with a spline.}
\label{fig: phot_all}
\end{figure*}

Our monitoring of SN 2014G lasted for $\sim1$ year, during which we gathered 407 photometric points distributed over 66 nights. The wavelength coverage is from the UV to the near-infrared (NIR) domains, and in the optical both Johnson/Cousins and Sloan Digital Sky Survey\footnote{http://www.sdss.org} (SDSS) filters were used. In addition, 18 optical spectra were taken, covering the $3000-10000$~\AA\ range (see Table \ref{tab: instr}). 2 UV spectra taken by \textit{Swift} + Ultraviolet/Optical Telescope\footnote{http://swift.gsfc.nasa.gov/} \cite[UVOT;][]{Roming 2005} are also available but they were omitted from this analysis because of contamination by nearby stars, which would have required the acquisition of further data to disentangle. Both the LC and the spectral evolution are well sampled, without significant temporal gaps.

All the CCD data have been corrected for overscan, bias and flatfields using standard procedures within {\tt IRAF}\footnote{http://iraf.noao.edu/}. Images from the Las Cumbres Observatory Global Telescope Network\footnote{http://lcogt.net/} \citep[LCOGT,][]{Brown 2013} were automatically ingested and reduced using the {\tt lcogtsnpipe} pipeline \citep{Valenti 2016}. For the photometric measurements, the {\tt SNOoPY}\footnote{Cappellaro, E. (2014). {\tt SNOoPY}: a package for SN photometry, http://sngroup.oapd.inaf.it/snoopy.html} package has been used, which allowed, for each exposure, to extract the magnitude of the SN with the point-spread-function (PSF) fitting technique \citep{Stetson 1987}. These magnitudes were then calibrated using the zero points and colour terms measured by reference to the magnitudes of field stars retrieved from the SDSS catalog (DR9). For the \textit{UBVRI} filters, we first converted the SDSS catalog magnitudes to Johnson/Cousins, following \cite{Chonis 2008}. The magnitudes of the Johnson/Cousins bands are thus given in Vega system, while the SDSS ones are calibrated in AB mag. The unfiltered images, considering the efficiency curve provided by the manufacturer of the camera, were calibrated to Cousins $R$-band. NIR data (\textit{JHK}) were reduced with a modified version of the external NOTCam package for {\tt IRAF}, including standard reduction steps of flat-field correction, sky background subtraction and stacking of the individual exposures for improved signal-to-noise ratio. The \textit{JHK} photometry was calibrated relatively to the magnitudes retrieved from the Two Micron All Sky Survey (2MASS) catalog\footnote{http://www.ipac.caltech.edu/2mass/}. Some precautions had to be taken with the \textit{Swift} data. \textit{Swift} is not equipped with a CCD but with a photon counter. For this reason, in the analysis we performed aperture photometry using the specific tools and parameters within the HEASARC\footnote{NASA's High Energy Astrophysics Science Archive Research Center.} software, and following \cite{Brown 2009}. For the optical spectra, the extractions were done using standard {\tt IRAF} routines. The spectra of comparison lamps and of standard stars acquired on the same night and with the same instrumental setting were used for the wavelength and flux calibrations, respectively. A cross-check of the flux calibration with the photometry (if available from the same night) and the removal of the telluric bands with the standard star were also applied.

The complete list of all the telescopes and instrumentations used to gather the data is reported in Table \ref{tab: instr}, while all photometric measurements are reported in Appendix \ref{Data} (Tables \ref{tab: ugriz} to \ref{tab: spec}).

\section[Photometry]{Photometry}\label{Photometry}
In Figure \ref{fig: phot_all}, the photometric evolution of SN 2014G in all bands is reported. \textit{Swift} UV observations (bottom left panel) started one day after discovery ($\sim3$~d after explosion) and lasted 23~d, during which the UV bands experienced a very steep and linear decline without showing any clear initial rise. The first NIR epoch, instead, was taken over one month after explosion and thus did not cover the rise time. In total, the NIR observations covered $\sim100$~d, but with only 5 epochs (bottom right panel). Focusing on the optical bands, a more complex behaviour than in UV and NIR bands is evident. The early dense coverage in \textit{UBVRIgri} (with first epoch being around $2-3$~d after discovery) allowed us to accurately track the rise to maximum. Fitting the curve with an order 3 polynomial, we inferred the peak epoch in different filters. These peaks were reached between 5.9$\pm$0.6~d in $U$-band (M$^{max}_{U}=-19.2$~mag) and 15.6$\pm$0.2~d in $i$-band (M$^{max}_{i}=-18.1$~mag) after the explosion, in agreement with what found by B16. In particular, the rise-time to maximum in $R$-band is 14.4$\pm$0.4~d (M$^{max}_{R}=-18.1$~mag), which, compared to the sample of \cite{Gall 2015}, supports the scenario of Type II-L SNe having longer rise-time and higher peak magnitudes than Type II-P. The LCs then declined linearly for $60-70$~d. Measuring the decline rates following the different prescriptions used in the literature, we found \mbox{$B_{100}=4.8$~mag}, \mbox{$V_{50}=1.5$~mag} and \mbox{$R_{50}=1.1$ mag}, thus confirming that SN 2014G comfortably matches the definitions used for Type II-L SNe (see slope limits reported in \mbox{Section \ref{sec: introduction}}). A steeper drop of $1.5-2$ magnitudes in $\sim10$~d occurred around day 80. This drop in particular is well sampled with very few comparable cases \citep{Valenti 2015,Valenti 2016,Yuan 2016}. The LC then appears to settle on a radioactive tail. The SN went behind the Sun after $\sim180$~d after explosion, and we were able to recover it at $\sim300$~d in four bands ($Vgri$).

\begin{figure}
\includegraphics[width=84mm]{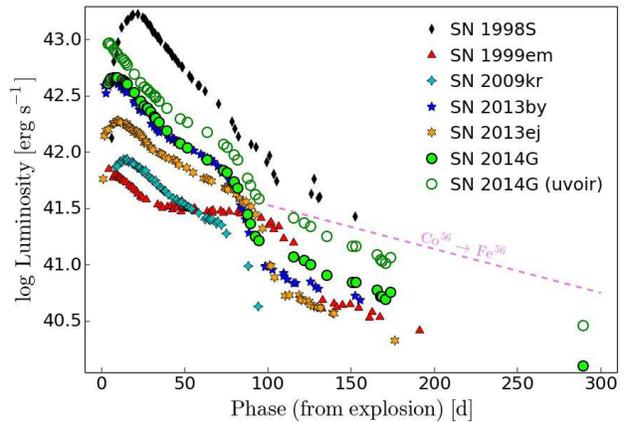}
\caption{Comparison of optical pseudo-bolometric LC of SN 2014G with those of SNe 1998S, 1999em, 2009kr, 2013by and 2013ej (see main text for references). For SNe 2014G, 2013by and 2013ej both Johnson/Cousins \textit{UBVRI} and SDSS \textit{ugri} filters were used, while just the former were used for the rest of the SNe. Note also that the U band was not available for SN 2009kr. For comparison we included also the \textit{uvoir} LC of SN 2014G, marked with green hollow circles. The dashed magenta line marks the slope that the LC would follow assuming that all the energy of the $^{56}$Co decay was absorbed by the ejecta.}
\label{fig: bol}
\end{figure}
\begin{figure}
\includegraphics[width=84mm]{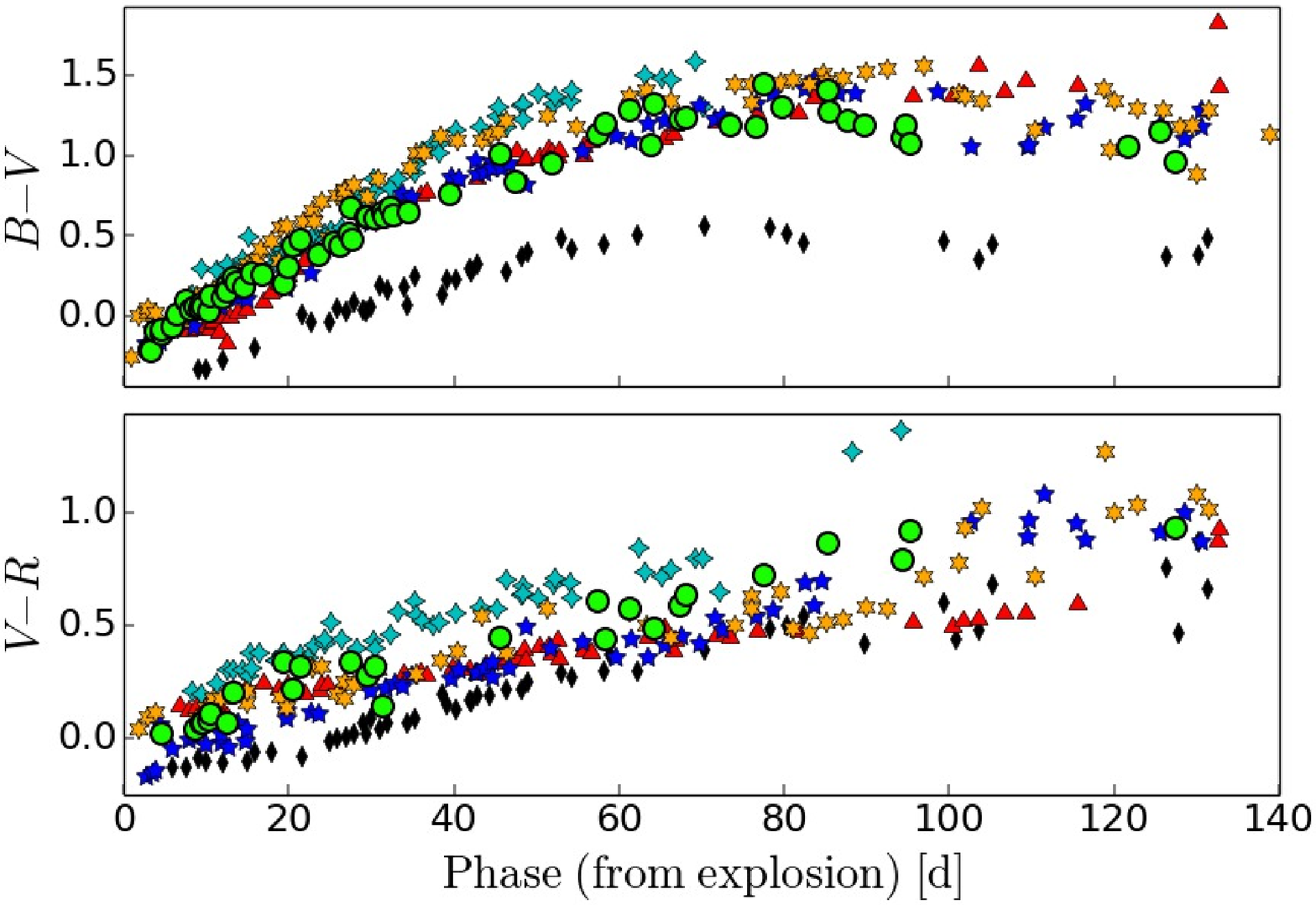}
\caption{$B-V$ and $V-R$ extinction-corrected colour evolution of SN 2014G compared to the other SNe considered in the text. The legend is the same as in Figure \ref{fig: bol}.}
\label{fig: col}
\end{figure}

\begin{figure*}
\includegraphics[width=168mm]{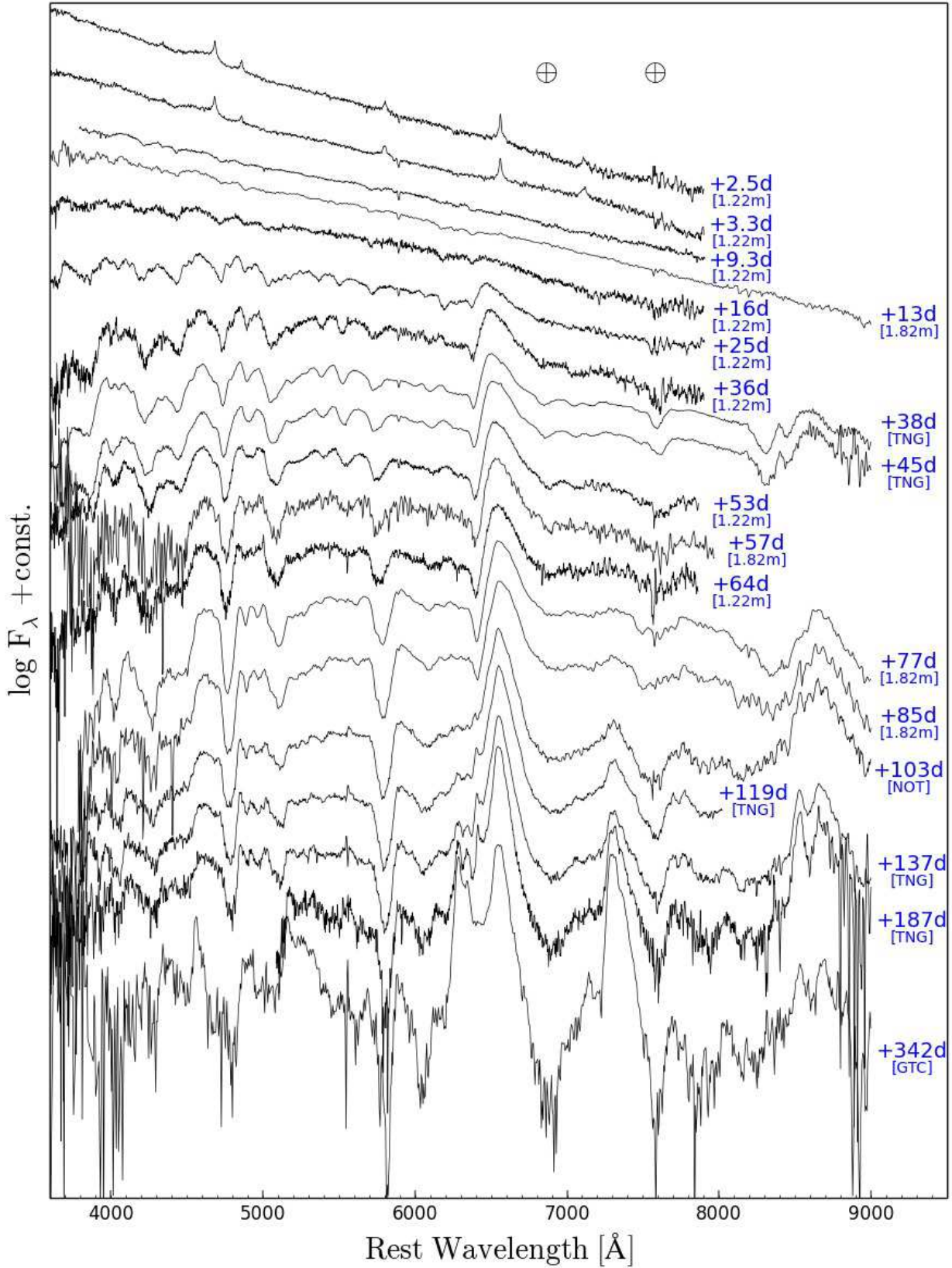}
\caption{Optical spectral evolution of SN 2014G. The spectra have been corrected for reddening and redshift, and shifted vertically for better display. On the right of each spectrum, the epoch and the telescope used are reported. The positions of major telluric absorption lines are marked with the $\bigoplus$ symbol.}
\label{fig: sp_ev}
\end{figure*}

\begin{figure*}
\includegraphics[width=168mm]{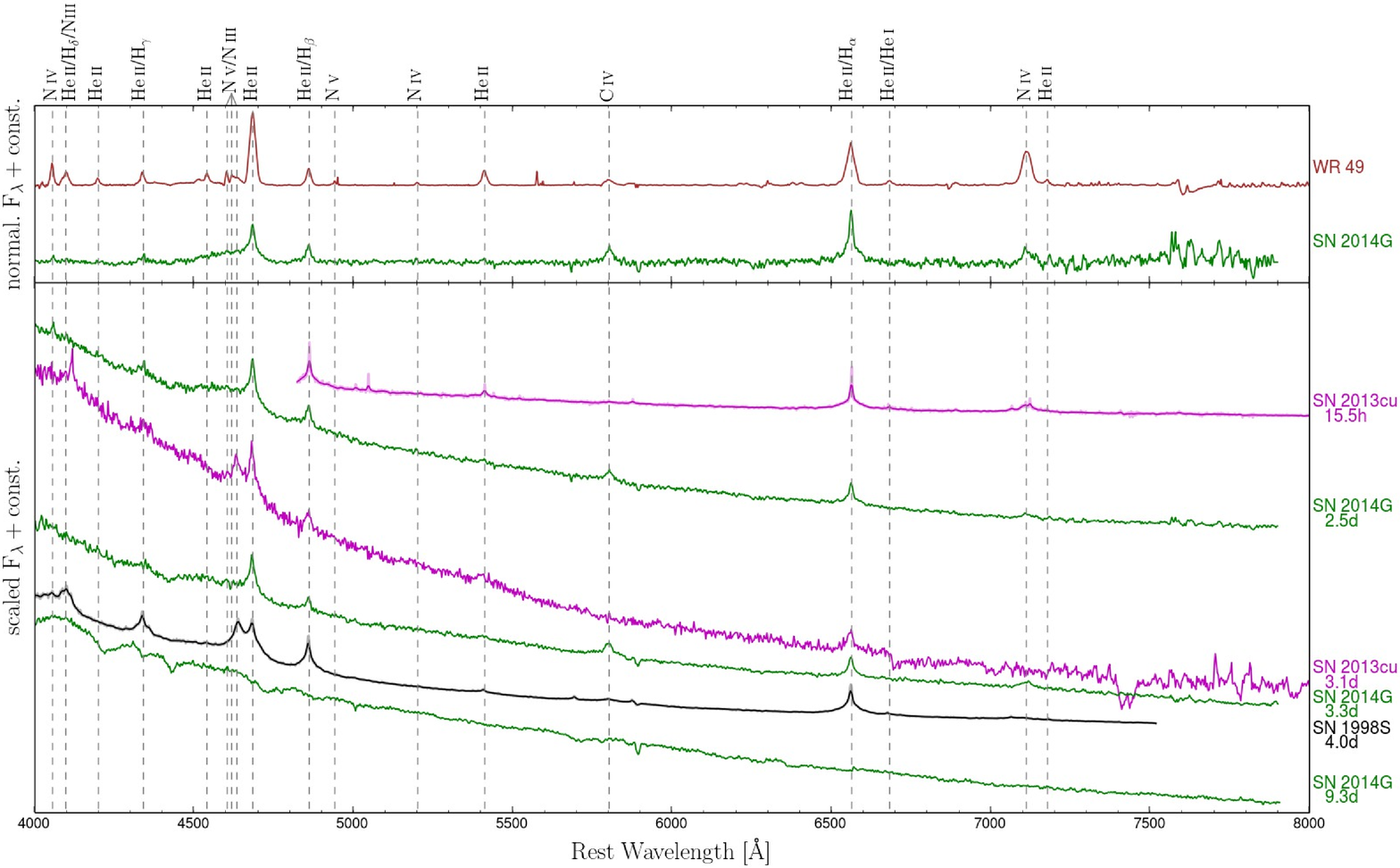}
\caption{\textit{Top panel}: Comparison between the normalised classification spectrum of SN 2014G and the WR star WN 49 \citep{Hamann 1995}. \textit{Bottom panel}: Comparison of early-phase spectra of SN 2014G with those of SNe 1998S \citep{Leonard 2000} and 2013cu \citep{Gal-Yam 2014}. The spectra of the two latter SNe were taken with a high resolution spectrograph, so a gaussian smoothing has been applied in order to match the resolution of the spectra of SN 2014G. Moreover, the spectra have been scaled for better comparison.}
\label{fig: early_sp}
\end{figure*}

 \begin{figure}
\includegraphics[width=84mm]{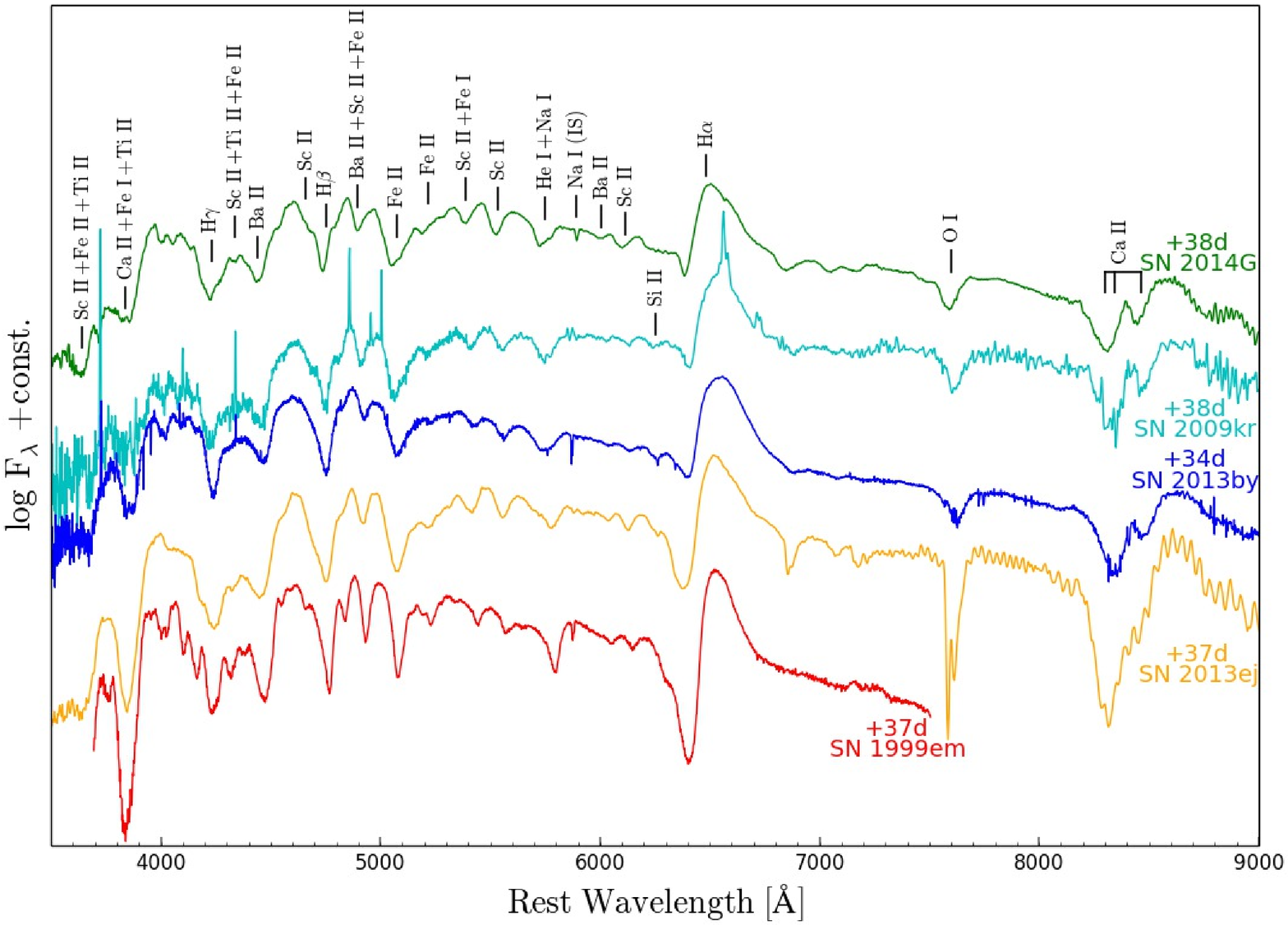}
\caption{Spectral comparison of SN 2014G with SNe 1999em \citep{Elmhamdi 2003}, 2009kr \citep{Elias-Rosa 2010}, 2013by \citep{Valenti 2015} and 2013ej \citep{Yuan 2016} at phase $34-38$~d. All the spectra are in rest frame and corrected for reddening.}
\label{fig: sp_comp}
\end{figure}

We then computed the \textit{uvoir} bolometric LC of SN 2014G, starting from the extinction-corrected fluxes at each epoch, and using the trapezoidal rule, assuming zero flux at the integration boundaries. All Johnson/Cousins \textit{UBVRI} and SDSS \textit{ugri} filters were used, along with the UV and NIR measurements. We also computed a pseudo-bolometric LC, using optical bands only. This was to allow meaningful comparisons with other SNe which do not have UV and NIR coverage. For this purpose, we selected some representative SNe from the literature, i.e. the Type II-L SNe 2009kr \citep{Elias-Rosa 2010}, 2013by \citep{Valenti 2015} and 2013ej \citep{Huang 2015,Valenti 2014,Yuan 2016}, the Type IIn SN 1998S \citep{Fassia 2000} and the archetypal Type II-P SN 1999em \citep{Elmhamdi 2003}. The comparison is shown in Figure \ref{fig: bol}. The shape of the LC of SN 2014G resembles those of SNe 2009kr and 2013ej. With respect to both of these though, SN 2014G shows a more rapid evolution, with shorter rise time, shorter duration of the photospheric phase and slightly steeper decline. The match with SN 2013by instead is striking, both in shape and luminosity, with the LCs of the two SNe matching almost perfectly. The only small difference is in the radioactive tail, which is more luminous in SN 2014G, indicating a greater amount of $^{56}$Ni synthesised. We notice however that SN 2014G declines faster with respect to the $^{56}$Co decay (dashed magenta line in Figure \ref{fig: bol}). One can argue that the missing flux could come from the NIR contribution, which is not optimally sampled by our data at these phases. However \cite{Inserra 2011} showed that at late phases the NIR contribution in Type II SNe is constant in time. Thus underestimating the NIR contribution in SN 2014G would translate into a solid shift of the tail but not into a change of the slope. We favour the idea that the steeper decline is due to a non-complete trapping of the $\gamma$-rays from radioactive decay (see Section \ref{sec: ni_mass}).

In Figure \ref{fig: col} we show the colour evolution $B-V$ and $V-R$ of SN 2014G, together with those of the other Type II SNe considered so far, all corrected for reddening. The behaviour of all SNe presented is quite similar, with a rapid increase of both colours, consistent with expectation from an expanding SN envelope. Only SN 1998S differs from the others, likely due to the contribution of CSM-ejecta interaction that characterised this transient \citep{Lentz 2001}. The excellent match of the colour evolution of SN 2014G with other Type II SNe supports the value of reddening adopted that was derived from the \ion{Na}{I} D (see Section \ref{NGC}).

\begin{figure*}
\includegraphics[width=168mm]{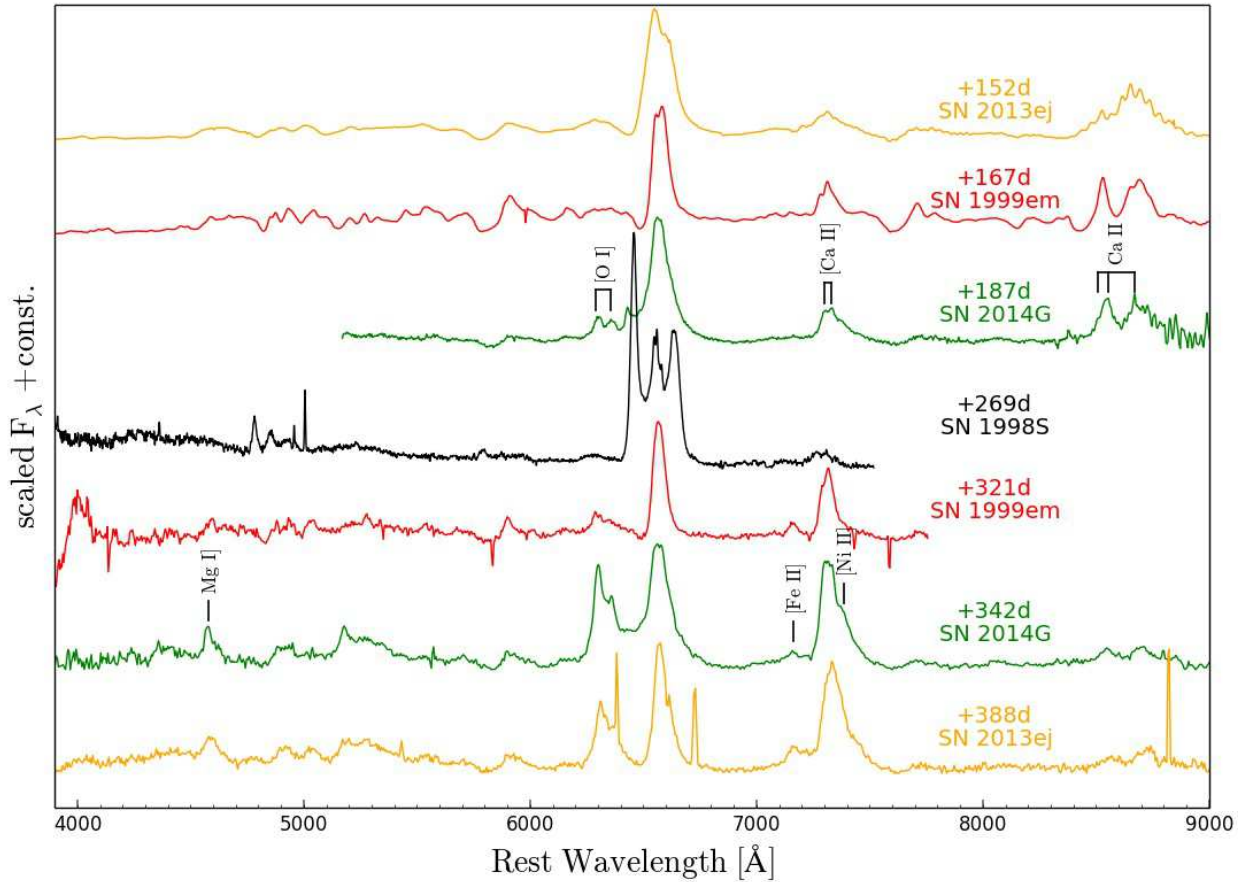}
\caption{Comparison of nebular spectra of SN 2014G with those of SNe 1998S \citep{Fransson 2005}, 1999em \citep{Elmhamdi 2003} and 2013ej \citep{Yuan 2016}. All the spectra are in rest frame and corrected for reddening, and have been scaled for better comparison. }
\label{fig: sp_neb_comp}
\end{figure*}

\section[Spectroscopy]{Spectroscopy}\label{Spectroscopy}
\subsection{Flash-ionised CSM lines}
Figure \ref{fig: sp_ev} shows the complete optical spectral evolution of SN 2014G from the classification spectrum to the nebular phase. The first five spectra show a blue continuum typical of Type II SNe at early phases. The first two, in particular, show emission lines that disappear after the first $\sim9-10$~d. These features have already been observed in a handful of early spectra of other CCSNe, such as SN 1998S \citep{Leonard 2000,Chugai 2001} or the Type IIb SN 2013cu \citep{Gal-Yam 2014}. The lines are emissions from highly ionised carbon and nitrogen, along with hydrogen and helium, and similar features are present also in the spectra of Wolf-Rayet (WR) winds. Indeed because of this coincidence, a WR progenitor star was originally proposed for SN 2013cu by \cite{Gal-Yam 2014}. However, this interpretation was disputed by \cite{Groh 2014} who modelled the emission lines and identified their origin in a slow dense wind or CSM surrounding the precursor star when it exploded. The wind velocity was estimated as $v_{\rm wind} \lesssim 100$~km\,s$^{-1}$ together with a mass loss rate of $\dot{\rm M}\simeq3\times10^{3}$M$_\odot$\,yr$^{-1}$. \cite{Groh 2014} proposed that this ruled out a WR-star progenitor, and favoured a luminous blue variable (LBV) or yellow hypergiant (YHG) progenitor with a wind that was enhanced in nitrogen and depleted in carbon (see Section \ref{subsec: progenitor}). We find the same lines in SN 2014G as in SN 2013cu, indicating that the surrounding CSM may have similar density and composition. A comparison among the early spectra of the above mentioned SNe is shown in Figure \ref{fig: early_sp} in addition to the WN5 type nitrogen-rich WR star WN 49 \citep{Hamann 1995}, to highlight the identification of these early-features. The main features of the spectra arise from H and \ion{He}{II} - the latter ion is seen as the line at $\lambda$4686. In addition, both SNe 1998S and 2013cu show a prominent peak at $\sim4630$~\AA\ which is probably a blend of \ion{N}{III} and \ion{N}{V}.
Several lines ($\lambda$4057, $\lambda$5201 and $\lambda$7113) are likely to be attributable to \ion{N}{IV}. A strong \ion{C}{IV} line at $\lambda$5803 is also evident in SN 2014G and this feature is missing in the other two SNe considered. Peculiar, narrow emission lines are also present in a high resolution spectrum of the Type IIb SN 1993J at 3~d after explosion \citep{Benetti 1994}. Together with \Ha and \ion{He}{II}, lines from [\ion{Fe}{X}] and [\ion{Fe}{XIV}] were identified, but these are not visible in SN 2014G. In summary, we find the same narrow emission lines of high ionisation species in SN 2014G as in SNe 1998S and 2013cu. They persist to at least 3~d after the estimated explosion date, and disappear by day 9 after explosion.

\subsection{Supernova lines}
All the narrow emission lines disappear in the spectrum at 9.3~d and more typical Type II features start to appear since the spectrum at 16~d. After this phase, a broad \Ha dominates the spectrum, with a P-Cygni profile characterised by an asymmetric emission and a shallow absorption. A couple of other Balmer lines (H$\beta$ and H$\gamma$) are identified, along with several lines of metal ions, like \ion{Ca}{II}, \ion{Fe}{II}, \ion{Sc}{II}, \ion{Ba}{II}, and \ion{Ti}{II}. A comparison with other SNe at these phases is shown in Figure \ref{fig: sp_comp} and the match with the spectra of SNe 2009kr and 2013by is very good. After $\sim80$~d the emission feature of [\ion{Ca}{II}] $\lambda\lambda$7291,7324 starts to become visible albeit weak at this epoch. The appearance of these lines is approximately coincident with the sudden drop in the LC. At a similar phase ($\sim100$~d) the \Ha absorption disappears and the emission line feature changes in structure - the interpretation of this will be discussed in Section \ref{sec: Halfa}.

At this point, the spectra show a gradual transition to the nebular phase with the [\ion{O}{I}] $\lambda\lambda$6300,6363 doublet becoming prominent. The forbidden Ca grows significantly in strength, with a red shoulder which can be attributed to [\ion{Ni}{II}] $\lambda$7378 from stable $^{58}$Ni \citep{Jerkstrand 2015a}, and a blue component of [\ion{Fe}{II}] $\lambda$7155 (\mbox{see Section \ref{sec: Ni/Fe}}). In the very last spectrum a prominent \ion{Mg}{I}] $\lambda$4571 is also present. A comparison with other SNe in the nebular phases is shown in Figure \ref{fig: sp_neb_comp}. At $\sim150-190$~d, distinct absorption components of H$\alpha$ are still present in SNe 2013ej and 1999em, whereas SN 2014G does not show any absorption. Instead at 103~d a narrow emission-like feature to the blue of the dominant H$\alpha$ emission appears (see Section \ref{sec: Halfa}). SN 1998S also shows a similar, but much stronger, narrow emission at approximately the same wavelength as in SN 2014G but in SN 1998S the H$\alpha$ line shows a complex blend with a triple peak structure. We notice also that the appearance of [\ion{O}{I}] occurs much earlier in SN 2014G than the other SNe shown here. Finally in the 342~d fully nebular spectrum H$\alpha$ and the [\ion{O}{I}] doublet appears to be connected by a distict ``bridge'' that links the two features (see Section \ref{sec: Halfa}). Apart from this anomaly, the nebular spectrum of SN 2014G is remarkably similar to that of SN 2013ej.

\begin{figure}
\includegraphics[width=84mm]{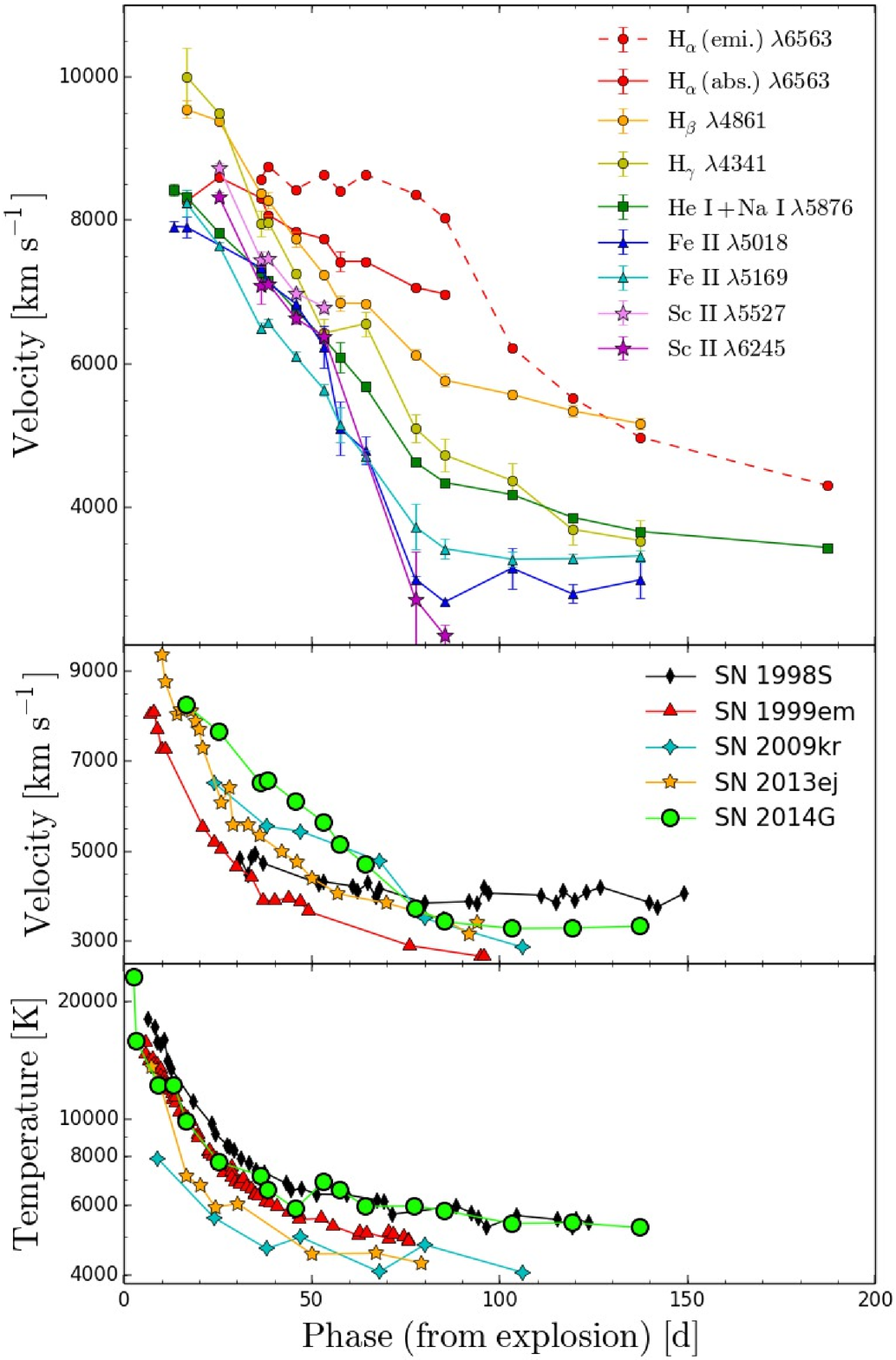}
\caption{\textit{Top panel}: velocity evolution of the Balmer lines and some prominent metal lines of SN 2014G. \textit{Middle panel}: comparison of the \ion{Fe}{II} $\lambda$5169 velocity of SN 2014G with that of the other SNe considered so far. \textit{Bottom panel}: temperature evolution of SN 2014G and comparison with other SNe. The values reported for SN 2009kr were measured from the spectra shown in Elias-Rosa et al. (2010), while for the other SNe we used the values published in the literature.}
\label{fig: lin_temp}
\end{figure}

\subsection{Line velocities and temperature evolution}
We measured the velocity of the important lines in each spectra throughout the first 150 days or for as long as feasible. The velocities were derived from the position of the minimum of their absorption features. For \Ha we 
measured also the FWHM of the emission component. Both the minima of the absorptions and FWHM of the H$\alpha$ emission were obtained from a fit with either a Gaussian, a Lorentzian or a low-order polynomial function, according to the best match with the shape of the feature. The errors were estimated with a Monte Carlo technique, varying the flux of each pixel according to a normally distributed random value having variance equal to the noise of the continuum. We did this procedure 100 times and then took the errors as the standard deviations of the fit parameters. 

The evolution of H$\alpha$, H$\beta$, H$\gamma$, \ion{He}{I}+\ion{Na}{I} $\lambda$5876, \ion{Fe}{II} $\lambda$5018 and $\lambda$5169, and \ion{Sc}{II} $\lambda$5527 and $\lambda$6245 are plotted in Figure \ref{fig: lin_temp}. We adopted the \ion{Fe}{II} $\lambda$5169 as a probe of the photospheric velocity \citep{Hamuy 2001} and compared its evolution with that of the other SNe considered in this paper (Figure \ref{fig: lin_temp}, middle panel). At early phases the ejecta velocity of SN 2014G looks higher than all the other SNe considered. Later on, however, it settles to a value of $\sim3300$ km~s$^{-1}$, in line with SNe 2009kr and 2013ej. 

From the spectra we estimated also the temperature evolution, obtained fitting the continuum of each spectra with a black-body function. The errors were calculated with the same Monte Carlo technique described above. The complete evolution is reported in Figure \ref{fig: lin_temp}, bottom panel, along with a comparison with the other SNe considered so far. The ejecta in SN 2014G appear to be hotter in comparison to the other similar SNe, with a late-time temperature comparable with that of SN 1998S. Moreover, we inferred the radius of the photosphere from the luminosity and the black-body temperature. The evolution of the first 3 points is well described by a parabola, with the vertex coincident with our estimate of the explosion date, strengthen the assumption we made in Section \ref{sec: introduction}.

\section[Discussion]{Results and analysis}\label{Discussion}

\begin{figure}
\includegraphics[width=84mm]{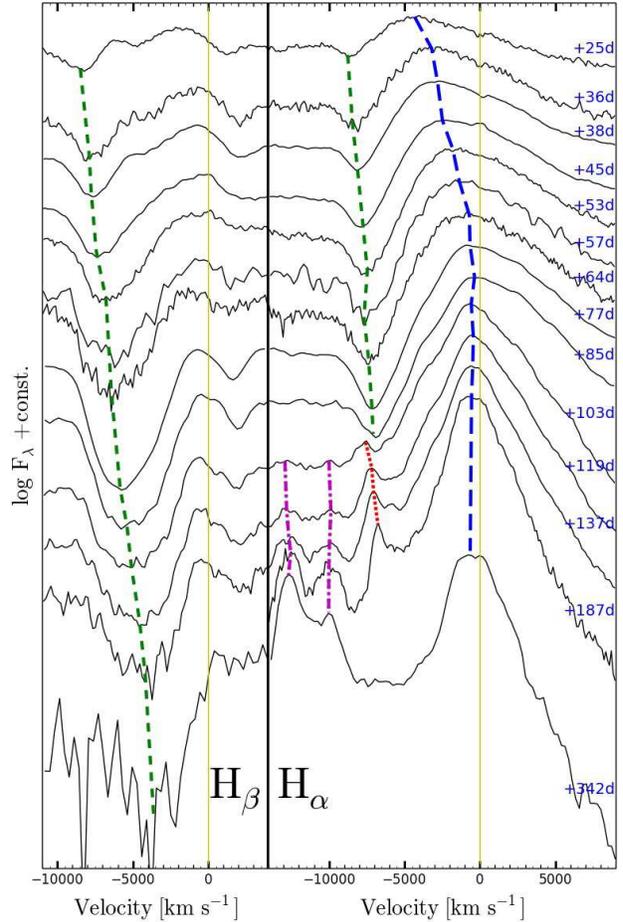}
\caption{\Ha (right panel) and H$\beta$ (left panel) profile history. The evolution of several features are marked: the maximum of the \Ha emission (long-dashed blue line); the \Ha and H$\beta$ minima (short-dashed green line); the [O I] doublet emission peaks (dot-dashed magenta line); the mysterious peak around 6400\AA\ (dotted red line). The rest frame zero velocity of both \Ha and H$\beta$ are also marked by solid yellow lines.}
\label{fig: Ha_ev}
\end{figure}

\subsection{H$\alpha$ evolution}\label{sec: Halfa}
Figure \ref{fig: Ha_ev} shows a zoom in of the \Ha and H$\beta$ profiles evolution (velocities are in the rest frame). The blue side of \Ha has been extended up to \mbox{$-14000$~km~s$^{-1}$} in order to include also the appearance of the [\ion{O}{I}] doublet $\lambda\lambda$6300,6364. 
The peak of the \Ha emission (blue dashed line) appears to be blue-shifted, by several thousands km~s$^{-1}$ at early phases, to only 500~km~s$^{-1}$ after $\sim80$ d. 
This is in contrast with the classical P-Cygni profile description, which predicts the emission to be at zero rest velocity. This behaviour has been already reported in many SNe \citep{Chevalier 1976,Taubenberger 2009}. This is thought to be a direct consequence of the steep density profile of the ejecta layers, which translates to more confined line emission and higher occultation of the receding part of the ejecta \citep{Anderson 2014b}. It is odd, however, that the emission never reaches the rest frame zero velocity, as one could expect at least at late phases.

The absorptions of both \Ha and H$\beta$, on the other hand, evolve steadily starting from about 9000 km~s$^{-1}$ to $6000-7000$\,km~s$^{-1}$ by day 100. Observations of other Type II-L SNe showed that this class tend to have less prominent \Ha P-Cygni absorptions then Type II-P\citep{Schlegel 1996,Gutierrez 2014}. This could be simply a consequence of a progenitor with less hydrogen in the ejected envelope, naively interpreted as less absorbing material along the line of sight. Models by \cite{Eastman 1996} alternatively suggested that this could be the consequence of a particularly steep density profile of the ejecta. \cite{Schlegel 1996} also proposed that weaker absorption can be the result of a smaller photosphere in comparison to the extent of the ejecta. In this scenario the column of absorbing material is narrow and weighted to lower velocities. 
The formation of a P-Cygni profile in SNe is not a trivial matter, and many factors can contribute simultaneously. Many of these however point towards more diluted ejecta, supporting the idea of Type II-L SNe having progenitors with less-massive and more spatially extended hydrogen envelopes than Type II-P SNe.

At $\sim100$~d a transition occurs in the blue part of H$\alpha$: the absorption feature disappears and is filled with a narrow emission at $\sim6395$~\AA. This peak is also present in the next three spectra and appears to have an evolution in velocity, marked in Figure \ref{fig: Ha_ev} by the red dotted line. No similar features are reported in the other SNe considered for comparison, with the possible exception of SN 1998S (see Figure \ref{fig: sp_neb_comp}). However, in the case of SN 1998S, there are two symmetric features around the H$\alpha$ zero velocity emission, together with similar features around other Balmer lines, interpreted as the result of the interaction of the ejecta with a disk-like structure \citep{Leonard 2000,Fransson 2005}. In the case of SN 2014G only a blue peak is visible, and is limited only to the \Ha line. We could not find a plausible identification for the source of this emission. Since its appearance is simultaneous with that of the [\ion{O}{I}] doublet, one might believe that these two features are related. However, the velocity evolutions are quite distinct, with the [\ion{O}{I}] remaining nearly steady (magenta dot-dashed line in Figure \ref{fig: sp_neb_comp}), while the mysterious peak moves redwards towards H$\alpha$. At this phase, the same velocity evolution is visible in H$\beta$, which might suggest a hydrogen origin for the feature. Assuming it is a high velocity feature of hydrogen, it evolves from $-7580$~km~s$^{-1}$ in the spectrum at 103~d to $-6755$~km~s$^{-1}$ at 187~d. High velocity hydrogen features have been identified before in other SNe spectra \citep{Inserra 2013}, although always in absorption and at much earlier phases. The feature in SN 2014G is somewhat reminiscent of the so-called Bochum event \citep{Hanuschik 1988}: the emergence of a blue and a red peak in the \Ha profile of SN 1987A after 20~d. The origin of the two peaks is independent, thus the absence of the red one in SN 2014G is not an issue. \cite{Hanuschik 1990} suggested that the blue peak is indeed an ``absorption deficit'' rather than an emission. This would be the result of a stratification of the hydrogen in the ejecta in 3 layers, the top and the bottom one being excited hydrogen, while the middle one would be constituted by ground state hydrogen with a low optical depth for H$\alpha$. This peculiar geometry would give rise to an emission-like feature at the velocity corresponding to the middle layer. However, even leaving aside the physical explanation for the complex structure to arise, this scenario does not comfortably explain our data, because a similar stratification would have created an anomalous feature also in H$\beta$, as in SN 1987A. Therefore we suggest that a Bochum event-like explanation does not quantitatively match what we see in the spectral evolution. 

\begin{figure}
\includegraphics[width=84mm]{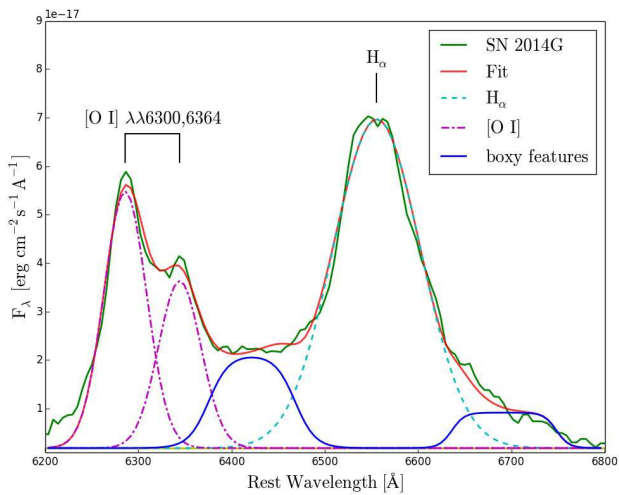}
\caption{Modelling of the $6200-6800$~\AA\ region of the SN 2014G spectrum at 342~d. We used gaussian profiles to reproduce \Ha and the [\ion{O}{I}] $\lambda\lambda$6300,6364 doublet and a boxy profile to reproduce the ``bridge'' in between.}
\label{fig: Ha_fit}
\end{figure}

In the spectrum at 342~d the mysterious peak disappeared, leaving place to a wider feature connecting \Ha and the [\ion{O}{I}] doublet. The elapsed time between the +187~d and the last +342~d spectra is too long to assert, with confidence, any direct evolution of the narrow emission at $\sim6400$~\AA\ to this ``bridge''. However the fact that we have two unusual features, at the same wavelength is peculiar and they may well be linked. We modelled the line emissions using gaussian profiles, and subtracting their contribution to the data, we obtain a flat-top feature corresponding to the ``bridge'' and also a noticeable excess of flux in the red wing of H$_\alpha$. Tentatively, we redid the fit with two boxy features symmetrically located with respect to the \Ha peak, reproducing nicely the overall emission profile in this region, as can be seen in Figure \ref{fig: Ha_fit}. The blue boxy profile is centred at $-5940$~km~s$^{-1}$ and it is $\sim4000$~km~s$^{-1}$ wide, while the red one is centred at 6270~km~s$^{-1}$ and with the same width. Note that we did not impose this condition, but this result came out from the fit. Flat-top profiles are usually interpreted as the ejecta interacting with a spherical shell of CSM \citep{Jefferey 1990,Chevalier 1994}, but it is quite unique to see two separate boxy profiles which are (roughly) symmetric with respect to the main emission. This feature could be attributed to a strongly bipolar geometry, namely from a jet-like flow of the ejecta interacting with a spherical CSM \citep{Smith 2012}. Or, vice versa, from a spherically symmetric ejecta interacting with an asymmetric CSM. The first scenario would create strong asymmetries in all the lines coming from the ejecta while here we see the boxy features only around H$\alpha$. Thus we favour the second scenario, in which the most external part of spherical ejecta starts to interact with an highly bipolar CSM. At this epoch (342~d) the interaction would thus be occurring only with the outer, hydrogen-rich part of the ejecta. Although one would expect to see similar features around all Balmer lines, the lower optical depth of other lines than \Ha could have prevented these features to be detectable. Highly asymmetric CSM structures are well known to exist around massive evolved stars \citep{Brandner 1997,Smith 2013,Smith 2016,Gvaramadze 2015}. The one surrounding the progenitor of SN 2014G would need to be composed of two polar lobes (like in the $\eta$-Car nebula), with a relatively narrow angle between their axis and the observer's point of view. Given a maximum velocity of the ejecta of $\sim10000$~km~s$^{-1}$ and the initial maximum velocity of the boxy feature of $\sim7600$~km~s$^{-1}$, taking the arccosine of the ratio of these two velocities we would infer an angle of $\sim40\degr$ between the observing axis and the axis of the polar lobes.

In this scenario, the emission-like feature which appeared in the spectra at $\sim6400$~\AA\ at $\sim100$~d could be interpreted as the beginning of the shock of the outer ejecta to the CSM. The redwards shift then, could be due to the reverse-shock travelling inwards through the ejecta. 
If this is the case, one may expect a red peak symmetrical to H$_\alpha$. However, the flux of the red boxy profile that we use to fit the line profile at late phases is less than half the blue one. So perhaps the red component at these phases is not bright enough to be detected, or there are radiative transfer effects across the CSM/ejecta interaction region that mask the receding material. Overall the fact that there are two broad components at either side of the \Ha emission peak, with approximately the same velocity, does suggest a bi-polar structure. 

\subsection{$^{56}$Ni mass}\label{sec: ni_mass}
Once all hydrogen in the envelope has recombined, the LC of a Type II SN settles onto the so-called radioactive tail. At this phase the energy source is the deposition of $\gamma$-rays and positrons originating from the decay chain $^{56}$Ni$\rightarrow^{56}$Co$\rightarrow^{56}$Fe. $^{56}$Ni has an e-folding time of 8.8~d thus the first decay is dominant in the very first part of the LC. However at these early phases the decay contribution is hidden by the hydrogen recombination power. $^{56}$Co, on the other hand, has a e-folding time of 111.4~d, thus the second decay of the chain is the one that shapes the LC when the recombination ends. The radioactive decay rate translates directly into a well defined slope of the LC (0.98~mag~100~d$^{-1}$), while the amount of $^{56}$Ni fixes the luminosity \citep[see][]{Cappellaro 1997}. As reported in \cite{Jerkstrand 2012}, the input energy during Co decay is
\begin{equation}\label{eq: 1}
L_0(t)=9.92 \times 10^{41} \frac{M_{^{56}\rm{Ni}}}{0.07 M_\odot}\left(\rm{e}^{-t/111.4} - \rm{e}^{-t/8.8}\right)~\rm{erg}~\rm{s}^{-1}~~,
\end{equation}
where $M_{^{56}\rm{Ni}}$ is the mass of $^{56}$Ni ejected during the explosion. In general in Type II SNe the $\gamma$-rays and positrons energy is fully trapped and thermalized, and the luminosity decline following the energy input, as shown with the dotted line in Figure \ref{fig: bol}. SN 2014G however declines faster, suggesting that the $\gamma$-ray trapping is incomplete.

\begin{figure}
\includegraphics[width=84mm]{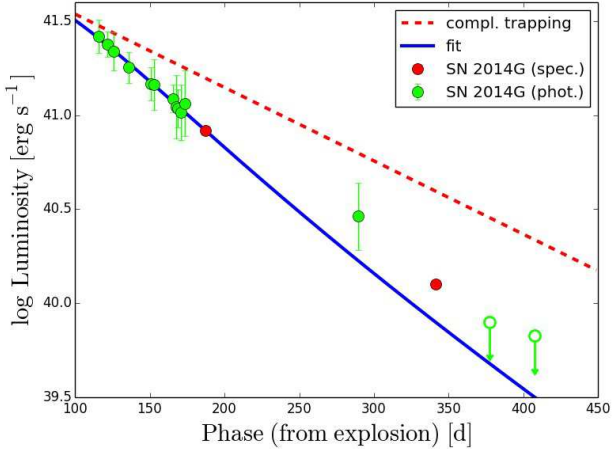}
\caption{Zoom in of the radioactive tail of SN 2014G. The green points correspond to the \textit{uvoir} bolometric LC, while the red ones are the integrated flux from the spectra. The red dashed line represents where the LC would be if it followed the $^{56}$Co decay in a full trapping regime. The blue solid line is the fit to the data with Equation \ref{eq: trapping}. The luminosities extrapolated from the spectra have high uncertainties, thus were not included in the fit.}
\label{fig: nichel}
\end{figure}

The problem of a non-complete trapping of the $\gamma$-rays has been analysed by \cite{Clocchiatti 1997} for the case of stripped-envelope SNe. They assume a simple model with spherical symmetry and homologous expansion, in which the $\gamma$-ray deposition is represented by a simple absorption process in radiative equilibrium. Then a simple expression can describe the luminosity:
\begin{equation}\label{eq: trapping}
L(t)=L_0(t)\times\left(1-\rm{e}^{-(\tau_\rmn{tr}/t)^2}\right)~~,
\end{equation}
where $L_0(t)$ comes from Equation \ref{eq: 1} and $\tau_\rmn{tr}$ is the full-trapping characteristic time-scale defined as 
\begin{equation}\label{eq: C}
\tau_\rmn{tr}=\left(D\kappa_\gamma\frac{M_\rmn{ej}^2}{E_k}\right)^\frac{1}{2}~~,
\end{equation}
where $M_\rmn{ej}$ is the total ejecta mass, $E_k$ is the kinetic energy, $\kappa_\gamma$ is the $\gamma$-ray opacity and $D$ is a constant which depends on the density profile (i.e. for a uniform density profile \mbox{$D=9/40\pi$}). The intent here is not to model the ejecta behaviour, rather to describe the radioactive contribution to the LC. For this purpose, Equation \ref{eq: trapping} is adequate and, as shown in Figure \ref{fig: nichel}, a good fit can be obtained with the observed \textit{uvoir} radioactive tail. Given the high uncertainties of the luminosities extrapolated from the spectra, we decided to not include them in the fit. A $^{56}$Ni mass of $0.059\pm0.003$~M$_\odot$ and $\tau_\rmn{tr}=162\pm10$~d is inferred \citep[for reference, for SN 1987A $\tau_\rmn{tr}=530$~d was estimated,][]{Jerkstrand 2011}. In Figure \ref{fig: nichel}, the dashed line shows the expected luminosity decline with the amount of $^{56}$Ni inferred and in case of complete trapping. Assuming a typical explosion energy of 1~foe, a uniform density profile of the ejecta, and a fiducial $\gamma$-ray opacity $\kappa_\gamma=0.03$~cm$^2$~g$^{-1}$ \citep{Colgate 1980}, from Equation \ref{eq: C} we can infer an indicative $M_\rmn{ej}\sim4.8$~M$_\odot$. Taking instead the 2.1~foe of kinetic energy estimated by B16 from their LC modelling, we infer $M_\rmn{ej}\sim7.0$~M$_\odot$, in agreement with what they obtained from the modelling. On the other hand, the $^{56}$Ni mass found is slightly larger than the amount found by B16, who reported three different measurements: 0.045 M$_\odot$ from the luminosity of radioactive tail, 0.055 M$_\odot$ from a comparison with SN 1987A and 0.052 M$_\odot$ from their LC modelling. In fact, not considering the incomplete trapping of the $\gamma$-rays, lead to underestimate the nickel mass. They considered a leakage of photons only in the LC modelling, but they might have underestimated the characteristic time-scale $\tau_\rmn{tr}$ by considering the early tail only. 

The $^{56}$Ni inferred for SN 2014G is within typical values for Type II SNe \citep{Hamuy 2003,Nadyozhin 2003,Sanders 2015,Valenti 2016}. Nevertheless, the presence of incomplete trapping is unusual for a Type II SN. 
It is clear from Equation \ref{eq: C} that the factors which can cause the incomplete trapping are essentially three: a low mass of the ejecta, a high kinetic energies or peculiar density profiles. \cite{Anderson 2014a} already found a significant number of Type II SNe with a $V$-band declining faster than the $^{56}$Co decay (with full trapping). They attributed this behaviour to low-mass and highly-diluted ejecta, which would be unable to completely trap the $\gamma$-rays. Since Type II-L SNe are supposed to arise from progenitors with rarefied envelopes \citep{Blinnikov 1993}, this could be a plausible scenario for SN 2014G. In addition to this, from Figure \ref{fig: lin_temp}, SN 2014G appears to have faster ejecta than the other SNe considered, possibly suggesting a particularly energetic event.
Equation \ref{eq: trapping} is appropriate if the Ni is in the centre of the ejecta. However, if there is strong mixing, and a considerable amount of Ni is spread out in the outer layers of the ejecta, then the $\gamma$-rays of the decay would have higher escape probability, and would then heat the surrounding material less efficiently, decreasing the luminosity. If this is the case, the ejecta mass could be higher than deduced from Equation \ref{eq: trapping}. One should note, however, that $^{56}$Ni
was strongly outmixed in SN 1987A, which still had full trapping for several hundred days. 

\subsection{Ni/Fe production ratio}\label{sec: Ni/Fe}
\begin{figure}
\includegraphics[width=84mm]{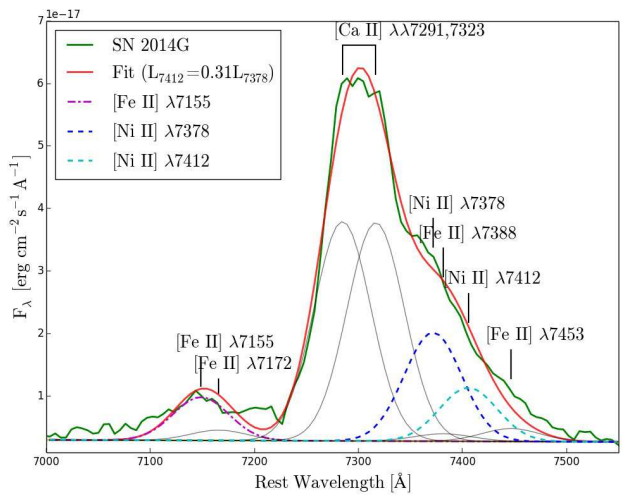}
\caption{GTC spectrum at 342~d between 7000 and 7600 \AA\ and the Gaussian fit described in Section \ref{sec: Ni/Fe} (red). The [\ion{Fe}{II}] $\lambda$7155 and the [\ion{Ni}{II}] $\lambda$7378 and $\lambda$7412 lines are also marked (see legend).}
\label{fig: Ca_fit}
\end{figure}

In the latest spectra, the [\ion{Ca}{II}] $\lambda\lambda$7291,7323 doublet showed a broad red shoulder. 
\cite{Jerkstrand 2015a} (hereafter referred to as J15) showed the case of \mbox{SN 2012ec} in which a prominent line was visible at this wavelength. This line was identified as an emission feature from stable $^{58}$Ni. The spectral models predict a distinct [\ion{Ni}{II}] $\lambda$7378 here, and in SN 2012ec the identification was made possible due to the relatively weak flux of the [\ion{Ca}{II}] doublet. Moreover the [\ion{Ni}{II}] line at 1.939~$\mu$m was also identified in a NIR spectrum. In the case of SN 2014G, however, the stronger [\ion{Ca}{II}] doublet and the higher blending makes the nickel identification less trivial. The feature could also be simply the result of asymmetries in the ejecta, such as we argued for H$\alpha$. However, in this case the excess is in the red, in contrast to \Ha where the stronger asymmetric component was shifted bluewards.
Nevertheless, we see that the \mbox{\ion{Mg}{I}] $\lambda$4571} line has also an asymmetric red shoulder.
So we attempted to fit the [\ion{Ca}{II}] feature using a doublet composed of two lines with the same velocity profile as observed in the single \mbox{\ion{Mg}{I}] $\lambda$4571} line. Alternatively, we fitted the whole feature with multiple gaussian profiles, including specifically lines at $\lambda$7378 and $\lambda$7412 representing [\ion{Ni}{II}].
The first method did not give a satisfactory fit and we then conclude that the red shoulder of the [Ca] doublet is better fit with an additional feature of [\ion{Ni}{II}] (cfr. Figure \ref{fig: Ca_fit}). 
This scenario is also physically supported by the fact that the material contributing to [\ion{Ca}{II}] emission is mostly situated in inner regions of the ejecta than the material emitting in \ion{Mg}{I}] \citep{Fransson 1989,Milisavljevic 2010}. Therefore differences in the shape of the profiles between the two ions are to be expected.


J15 presented a new analytic method to determine the Ni/Fe ratio in nebular spectra of SNe in the $7100-7400$ \AA\ region. The physical regimes for which the method is valid was confirmed by inspecting the conditions in forward spectral simulation models. Here, we apply the analytic method to SN 2014G, despite the quantitative measurement of [\ion{Ni}{II}] $\lambda$7378 is not as easy as in the case of SN 2012ec, where the line is resolved.
In the above mentioned spectral region, there are 8 prominent emissions: [\ion{Ca}{II}] $\lambda\lambda$7291,7323, [\ion{Fe}{II}] $\lambda$7155, $\lambda$7172, $\lambda$7388 and $\lambda$7453, [\ion{Ni}{II}] $\lambda$7378 and $\lambda$7412 (see Figure \ref{fig: Ca_fit}). Following J15, we fixed the strength ratio between the lines of the same species: the Fe lines $\lambda$7155 and $\lambda$7453 come from the same atomic level, and thus their luminosity ratio is constant with L$_{7453}=0.31$L$_{7155}$. Also the other two Fe lines come from the same level, thus we imposed L$_{7388}=0.74$L$_{7172}$. We could also fix the ratio between the [\ion{Fe}{II}] $\lambda$7155, $\lambda$7172, despite coming from two different levels. This ratio depends only weakly on the temperature, and following the J15 model we assumed L$_{7172}=0.24$L$_{7155}$. J15 coupled each line to [\ion{Fe}{II}] $\lambda$7155, but overall the line ratios among the iron lines were the same as ours. On the other hand, the two Ni lines come from two different levels and their luminosity ratio depends non-negligibly on temperature. However J15, based on their model with $\rmn{T}=3180$~K, fixed L$_{7412} = 0.31$L$_{7378}$. We employed the same approach, but in appendix \ref{sec: Ni/Ni} we investigated how to relax this constraint.
We used simple gaussians as fitting profiles, forcing all the lines to have the same velocity $\Delta v$ (i.e. the same FWHM) but allowing also a rigid shift $\Delta\lambda$ of the line centroids (however keeping fixed the relative position of each line). In total we had 5 free parameters and our best fit with this set-up is shown in Figure \ref{fig: Ca_fit}. The values inferred from this fit were \mbox{L$_{7155} = 4.76
\times10^{-16}$~erg s$^{-1}$}, \mbox{L$_{7291}=2.37
\times10^{-15}$~erg s$^{-1}$}, \mbox{L$_{7378}=1.36
\times10^{-15}$~erg s$^{-1}$}, \mbox{$\Delta\lambda=-6.7
$~\AA}, \mbox{$\Delta v=2605
$~km}. This gave the final result of \mbox{L$_{7378}$/L$_{7155}=2.9\pm0.2$}.

The iron and nickel content can be inferred by the ratio of the luminosity of the [\ion{Fe}{II}] $\lambda$7155 and [\ion{Ni}{II}] $\lambda$7378 lines following the relation
\begin{equation}\label{eq: Ni/Fe}
\frac{L_{7378}}{L_{7155}}=4.9\left(\frac{n_\rmn{Ni}}{n_\rmn{Fe}}\right)\rm{e}^{0.28\rmn{eV}/kT}~,
\end{equation}
where $n_\rmn{Ni}$ and $n_\rmn{Fe}$ are the number densities of \ion{Ni}{II} and \ion{Fe}{II}, $k$ is the Boltzmann constant and $T$ the temperature (see J15 for the origin of the constants). The temperature can be constrained from the luminosity of one Fe line and the total Fe mass, assuming local thermodynamic equilibrium (LTE). The iron production is dominated by the $^{56}$Ni$\rightarrow^{56}$Co$\rightarrow^{56}$Fe chain. At these late phases, the Ni has all decayed into Co, and 95 per cent of it has already become Fe, thus the iron mass is constrained by the $^{56}$Ni mass we measured in the previous Section (0.059~M$_\odot$). Then assuming that most of iron is in the form of \ion{Fe}{II} (J15 find it to be around 90 per cent at 370~d), we can write (see J15 for the derivation of the constants)
\begin{equation}
\frac{L_{7155}}{M(^{56}\rmn{Ni})}=\frac{8.67\times10^{43}}{15 + 0.006T}\rm{e}^{-1.96\rmn{eV}/kT}~\rmn{erg~s^{-1}M}^{-1}_\odot~.
\end{equation}
This function is really steep, which translates in the temperature varying little for changes in the $L_{7155}/M(^{56}$Ni) ratio (see Figure 6 in J15). From this equation, we obtained $T=2701^{+44}_{-47}$~K. Then putting this value in Equation \ref{eq: Ni/Fe} we infer $n_\rmn{Ni}/n_\rmn{Fe} = 0.18\pm0.02$. This is very similar to 0.19 found by J15 for SN 2012ec. As they pointed out, these values are considerably higher than the 0.06 solar abundance ratio. This is the fourth SN with a significantly supersolar Ni/Fe production \citep{MacAlpine 1989,Maeda 2007,Mazzali 2007}. Several others show solar or subsolar (J15), so there seems to be significant diversity. Primordial Fe and Ni contamination could have contributed to the line flux we measured from the fit. As deduced by J15, potentially this contamination could have had underestimated the Ni/Fe ratio in the iron zone, however less than $\sim1/3$. 

\cite{Jerkstrand 2015b} demonstrated through nucleosynthesis simulations that a high Ni/Fe ratio, like the one found for SN 2014G, imply burning and ejection of the silicon-layer material in the progenitor, with neutron excess $\eta\sim6\times 10^{-3}$. Such a process is most easily achieved in lower mass progenitors (M$_\rmn{ZAMS}<13$~M$_\odot$) exploding with a delay time of less than 1 second. However, strongly asymmetric explosions may also achieve relatively high Ni/Fe ratios in more massive progenitors.


\subsection{The progenitor}\label{subsec: progenitor}
\begin{figure*}
\includegraphics[width=168mm]{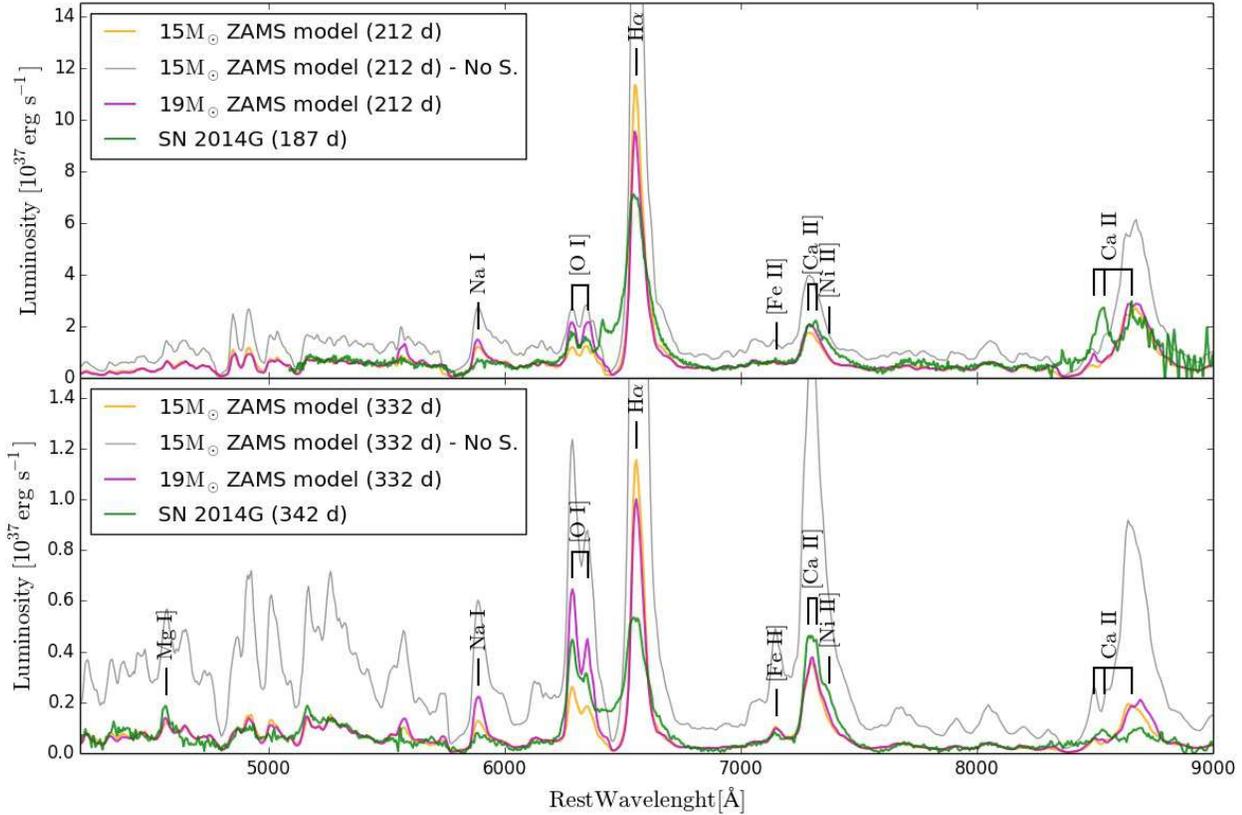}
\caption{Comparison of the nebular spectra of SN 2014G at 187~d (top panel) and 342~d (bottom panel) with the model spectra from Jerkstrand et al. (2014). We consider the models with progenitor of M$_\rmn{ZAMS}$ of 15 and 19 M$_\odot$ at phase 212 and 332~d. The models were scaled to same distance and same $^{56}$Ni mass as SN 2014G. Also the incomplete trapping was considered, and as reference we included in grey the model of M$_\rmn{ZAMS}=15$ without the scaling. Note in particular that the [Ni II] $\lambda$7378 line looks underproduced in the models, as they have close to solar Ni/Fe ratio (see Section \ref{sec: Ni/Fe}).}
\label{fig: ajneb}
\end{figure*}
The high ionisation lines of \ion{He}{II}/\ion{N}{IV}/\ion{C}{IV} seen in the early spectra of the Type IIb SN 2013cu addressed \cite{Gal-Yam 2014} to infer a WR progenitor for that transient. However, \cite{Groh 2014} and \cite{Grafener 2016} modelling the early spectra of SN 2013cu found that chemical composition, mass loss-rate and wind velocity are consistent with the properties of a LBV, a YHG or an extreme RSG. 
Also, \cite{Shivvers 2015} studied the early spectra of SN 1998S and found a slow wind consistent with that of a RSG progenitor. A similar interpretation was given by \cite{Smith 2015}, and they conclude that the early-time WR-like spectrum has little to do with the spectral type of the progenitor before explosion. In fact N-rich nebulae can also be found around evolved massive stars occupying the upper Hertzsprung-Russell diagram \citep{Smith 1997,Lamers 2001, Smith Morse 2004}.

A sample of spectroscopy from young Type II SNe by \cite{Khazov 2016} has recently been published showing that about 14 per cent of all Type II SNe which have spectra available within 10 days of explosion have these high ionisation features present. They also suggest that the SNe with these features tend to have brighter peak magnitudes than average for Type II SNe and SN 2014G would fit well with that picture. Although we present one object, in comparison to the sample of 12 of \cite{Khazov 2016}, our comprehensive dataset allows us to track this SN into its latest stages to probe the nucleosynthesis within the progenitor and estimate its mass. 

From the modelling of the LC, B16 inferred a progenitor radius of 630~R$_\odot$. 
Therefore the progenitor of SN 2014G was likely to be an extended supergiant star, with at least part of its H-rich envelope left at the time of explosion. Current theories for the progenitors of Type II-L involve a more extended less-massive hydrogen envelope with respect to Type II-P SNe progenitors \citep{Blinnikov 1993}, and many aspects of SN 2014G favour this scenario. First of all, the LC in the photospheric phase points towards a more diluted envelope, unable to sustain a flat luminosity plateau. Secondly, the faster decline of the radioactive tail with respect to complete trapping of $\gamma$-rays from $^{56}$Co decay suggests less-dense ejecta. This is also consistent with the blue-shift of the \Ha emission of SN 2014G, which shows a blue-shifted emission up to 5000 km~s$^{-1}$ at 25~d and settles close to zero velocity already around 65~d. In fact \cite{Anderson 2014b} showed that the blue-shift should be higher in the early spectra for objects with more extended envelopes, and it should also evolve towards rest frame zero velocity much faster than for more compact objects. 

We finally tried to constraint the mass of the progenitor. \citet[][the latter hereafter referred to as J14]{Jerkstrand 2012,Jerkstrand 2014} showed how the spectral modelling of the nebular phase can actually be used to constrain the M$_\rmn{ZAMS}$ of the progenitors of Type II SNe \citep[see also][]{Dessart 2010,Dessart 2011}. They find, in particular, that the flux of the [\ion{O}{I}] $\lambda\lambda$6300,6364, \ion{Na}{I} $\lambda\lambda$5890,5896 and \ion{Mg}{I}] $\lambda$4571 lines show a strong correlation with the progenitor mass, with the oxygen doublet being the most important as a diagnostic of the core mass. We took their synthetic spectra as a comparison for the nebular spectra of SN 2014G at 187 and 342~d, seeking the best match with the [\ion{O}{I}] lines flux. We point out that, as it is possible to see from Figure \ref{fig: Ha_fit}, the ``bridge'' feature between the oxygen doublet and \Ha does not significantly affect the oxygen flux, allowing a direct comparison between the data and the models.

In order to perform a correct comparison, the synthetic spectra had to be scaled to the same distance of SN 2014G and to the same amount of $^{56}$Ni. Moreover, having $\tau_\rmn{tr}$ significantly higher than the one inferred for 2014G (470 and 530~d for the model of 15 and 19 M$_\odot$ M$_\rmn{ZAMS}$ progenitor, respectively), we rescaled the models using Equation \ref{eq: trapping}, to take account of the missing flux due to the higher leakage of $\gamma$-rays. Finally we also corrected for the small difference in phases, reducing the cobalt contribution according to how much has decayed in the time between the epoch of the model and the epoch of the observed spectrum. The comparison is reported in Figure \ref{fig: ajneb}. The level of the quasi-continuum of the models matches that of the spectra of SN 2014G, suggesting that the opacity of the ejecta to $\gamma$-rays is in fact described sufficiently well by Equation \ref{eq: trapping}. From the comparison it is clear that the [\ion{O}{I}] $\lambda\lambda$6300,6364 doublet of SN 2014G sits between the model with a progenitor with M$_\rmn{ZAMS}$ of 15 and 19 M$_\odot$. Therefore, we infer a progenitor with M$_\rmn{ZAMS}\simeq17\pm2$~M$_\odot$. A particularly low hydrogen flux looks also evident from the comparison with the synthetic spectra. This again favours the scenario of Type II-L SNe having progenitors with a reduced hydrogen envelope at the moment of explosion \citep{Popov 1993}. This is one of the first times that this method has been applied on a Type II-L SN \citep{Valenti 2016,Yuan 2016}, and this is also arguably one of the most massive ZAMS progenitor for a Type IIP/L so far \citep{Smartt 2015}, favouring the scenario of Type II-L SNe arising from more massive ZAMS stars than Type II-P \citep{Elias-Rosa 2010,Elias-Rosa 2011}.


Small differences between the observed spectra and the models are present, in particular for other tracers of M$_\rmn{ZAMS}$, like \ion{Mg}{I}] $\lambda$4571 and \ion{Na}{I} D. Modelling of the Mg and Na lines is, however, more complex than the modelling of [\ion{O}{I}] $\lambda\lambda$6300,6364, with increased sensitivity to density and ionisation conditions. Both elements have only small fractions in the neutral state, and the exact value governs whether there is a cooling contribution or just a recombination contribution \citep[see][]{Jerkstrand 2015c}. Moreover, sodium shows a somewhat erratic growth with M$_\rmn{ZAMS}$ \citep[e.g.][]{Woosley 1995}. A significant part of the \ion{Na}{I} line is caused by scattering in hydrogen-rich gas, and a reduced hydrogen-zone mass in SN 2014G may be responsible the weak line. The oxygen on the other hand is mainly neutral, and thus its cooling emission less sensitive to ionisation ratios. Thus, in lack of ad-hoc modelling, focusing only on the forbidden oxygen doublet is the preferred way to proceed.

As previously mentioned, we addressed the issue of the non-complete trapping observed in SN 2014G by down-scaling the models with Equation \ref{eq: trapping}. However, since the M$_\rmn{ZAMS}$ estimate is based on line comparison, one should focus on how the incomplete trapping is influencing the line formation, which is not trivial. The mass loss in the \cite{Woosley 2007} stellar evolution models used in J14 is computed with standard recipes at solar metallicity, and gives only minor mass loss for M$_\rmn{ZAMS} < 20$~M$_\odot$. In the J14 models, the ejecta have a defined morphology, i.e. a core where $^{56}$Ni, Si/S, O and part of the He and H zones are macroscopically mixed, and an unmixed envelope. The mixing structure is guided by the morphologies obtained in multi-dimensional simulations. Thus the incomplete trapping of the $\gamma$-rays can be due to the following scenarios (see Section \ref{sec: ni_mass}):
\begin{itemize}
\item A less massive hydrogen envelope. The envelope itself does not influence directly the line formation of core elements like oxygen. If we then assume a core with unchanged properties, the core of SN 2014G would still be similar to the models, and the \ion{O}{I} luminosity should not be scaled with any trapping function. The spectra would then be overall dim because of the lost of the hydrogen zone deposition. But if the core structure has changed due to the hydrogen envelope loss (which is likely), weaker in-mixing of oxygen leads to dimmer oxygen lines. Thus, the models are too bright and need to be scaled down.
\item High kinetic energies. The ejecta are expanding faster due to higher energies, and the trapping is weaker. Thus the models are too bright in all zones and a down-scaling is appropriate.
\item Higher $^{56}$Ni mixing. The whole ejecta are illuminated in a more dilute $\gamma$-field and $\gamma$-rays escape starts earlier. Oxygen is heated less efficiently and thus a uniform down-scaling is appropriate. 
\end{itemize}
We favour the idea that mixing played an important role, as it is also shown by the presence of carbon an nitrogen in the early spectra, and thus that a down-scaling of the models is necessary. In Figure \ref{fig: ajneb} we showed, in grey, the 15 M$_\odot$ ZAMS model scaled only to the $^{56}$Ni mass of SN 2014G, but with the $\gamma$-ray deposition, including escape, computed in the models. The difference in flux in all wavelength between the model and the observed spectrum is evident, again suggesting a down-scaling of the models in order to perform a satisfying comparison. Everything considered, the comparison reported in Figure \ref{fig: ajneb} should be correct, and the estimate of a relatively massive ZAMS progenitor for SN 2014G consistent.

B16 modelled the LC of SN 2014G with a semi-analytical approach, inferring an ejecta mass of only 7~M$_\odot$, plus 2~M$_\odot$ for the compact remnant. Assuming a fiducial error of 1~M$_\odot$ on their estimate, the $M_{\rm ZAMS} \simeq 17\pm2$M$_\odot$ estimated by us might indicate a mass loss between 5 and 11~M$_\odot$, from ZAMS to the explosion. The ``classical'' empirical mass loss rate for RSGs given by \cite{de Jager 1988} predicts
that progenitors up to $M_{\rm ZAMS} \simeq 25$M$_\odot$ should retain enough mass to produce a Type II-P SN. More recent works, however, showed how RSG stars could have experience much higher mass loss in their main sequence phase than previously thought, with several observations supporting this statement \citep{Smith 2004,van Loon 2005,Davies 2008}. The mechanism at the origin of this enhanced mass loss is not clear, with pulsation-driven superwinds \citep{Yoon 2010a} or super-Eddington luminosities in the outer layers of the star \citep{Ekstrom 2012} as two possible physical reasons. In the former scenario, \cite{Yoon 2010a} find pulsations to occur from a M$_\rmn{ZAMS}$ of 17~M$_\odot$, and they also infer that a 20~M$_\odot$ ZAMS progenitor would end its life as a yellow supergiant (YSG) of 6.1~M$_\odot$, 0.5~M$_\odot$ of which is hydrogen. \cite{Ekstrom 2012} instead worked on models with enhanced mass loss due to the luminosity in the external layers exceeding the Eddington limit. From their rotating models of ZAMS stars of 15 and 20~M$_\odot$ they inferred a mass of $\sim11$ and 7~M$_\odot$ respectively at the end of carbon burning phase. According to these simulations then, the 17~M$_\odot$ M$_\rmn{ZAMS}$ progenitor of SN 2014G inferred in this work could have reasonably led to a pre-explosion progenitor of only 9~M$_\odot$, as the one inferred by B16. Without direct observations of the progenitor we cannot discriminate between a RSG or a YSG. Several Type II SNe have observational evidence for YSG progenitors \citep{Maund 2004,Maund 2011,Elias-Rosa 2009,Elias-Rosa 2010,Fraser 2010}. \cite{Georgy 2012} showed how these YSG progenitors could actually be the end of the evolution of RSG stars with enhanced mass loss, which may qualitatively match our observations for SN 2014G. We also have to take into consideration that a companion star in a close-binary system could also have been the culprit for the mass loss, through a Roche lobe overflow. \cite{Yoon 2010b} showed how the mass loss in close binary for a ZAMS star of 17~M$_\odot$ could have been much more extreme, leaving a star of only $\sim4$~M$_\odot$. 
A fine tuning of distance$-$mass ratio between the two stars could easily be able to reproduce the inferred mass loss \citep{Podsiadlowski 1993,Claeys 2011, Eldridge 2011}. In Section \ref{sec: Halfa} we suggest that an highly asymmetric CSM was the origin of the atypical \Ha profile, which could be consistent with mass loss events in binary systems. Both of these scenarios put strong constraints on current explosion theories, and the growing number of SNe with measured explosive burning products from nebular spectra gives hope for progress in understanding how CCSNe explode.

\section[Summary and conclusions]{Summary and conclusions}
We presented \textit{uvoir} photometry and optical spectroscopy for the Type II-L SN 2014G up to 342~d after explosion. Our detailed dataset allowed us to investigate many different aspects of its evolution from very early after explosion to 
observations deep into the nebular phase. The early spectra show narrow emission lines that disappear by day 9. These
high ionisation lines have been seen in other SNe of Types II-P/L and IIb and suggest that the CSM surrounding the 
star is photo-ionised by UV emission at shock breakout. This study shows one example in which they exist in Type II-L SNe, and persist to 
at least 3~d after explosion and disappear by about 10 days. One should be careful with classifications that 
are based on such early spectra, as a IIn classification would typically result. 
The LC then evolved like a canonical Type II-L, with a linear decay of the LC lasting until $\sim80$~d after explosion. The radioactive tail appeared to fall more rapidly than the $^{56}$Co decay with full trapping. We interpreted this as a leakage of $\gamma$-rays, possibly due to more diluted external layers of the ejecta and high levels of mixing. Taking into account the missing flux, we were able to infer a $^{56}$Ni mass of $0.059\pm0.003$~M$_\odot$. 

We further presented extensive late-time spectral coverage of SN 2014G offering further insights into the progenitor and explosion. \Ha became optically thin already at 100~d, the spectra showed an early [\ion{O}{I}] doublet and an intriguing narrow emission between these two lines, which evolved later into a wider and flat-topped boxy profile. We interpreted this feature as the interaction of the outer hydrogen rich ejecta with a strongly asymmetric CSM. We computed a combined line
profile of H$\alpha$ in the 342~d spectrum with the sums of multiple individual velocity components. 
We showed some evidence that there are two broad, boxy features roughly symmetrically distributed at either side of zero velocity
that would explain the complex \Ha region. 
This symmetry led us to infer a strongly bipolar CSM geometry, with one of the lobes oriented toward us, at angle of $\sim40\degr$ with the observer line of sight. 
In the last spectrum the [\ion{Ca}{II}] $\lambda\lambda$7291,7323 doublet shows a distinct flux excess on the red side. We found this flux excess consistent with the presence of emission lines, which we identified as [\ion{Ni}{II}]. Applying a semi-analytical line formation method, we were able to infer a Ni/Fe production ratio, obtaining a value of 0.18, 3 times higher than the solar ratio.

Finally we investigated and discussed the nature of the progenitor. We compared the nebular spectra with the models of J14, focusing on the flux of the [\ion{O}{I}] $\lambda6300,6364$ doublet. From this comparison, we inferred a progenitor with a ZAMS mass of 17~M$_\odot$. From the LC modelling, B16 inferred a progenitor with an ejecta mass of 9~M$_\odot$, thus our result implies an extensive mass loss of the progenitor during its life. Therefore, 
the progenitor of SN 2014G was likely to be a RSG or YSG which experienced extensive mass loss. Traces of the mass loss are the photo-ionised metal-rich CSM inferable from the early spectra and the hints of interaction visible in the late ones. Such extensive mass loss left the progenitor with a low-mass and highly diluted envelope. Overall, SN 2014G is supporting the current theories of Type II-L SNe arising from massive hydrogen-depleted stars. 

\section*{Acknowledgments}
GT, SB, EC, NE-R, AH, AP, LT, and MT are partially supported by the PRIN-INAF 2014 with the project Transient Universe: unveiling new types of stellar explosions with PESSTO. NE-R acknowledges the support from the European Union Seventh Framework Programme (FP7/2007-2013) under grant agreement no. 267251 ``Astronomy Fellowships in Italy'' (AstroFIt). The research leading to these results has received funding from the European Research Council under the European Union's Seventh Framework Programme (FP7/2007-2013)/ERC Grant agreement n$^{\rm o}$ [291222] (PI : S. J. Smartt) and STFC grants ST/L000709/1. AMG acknowledges financial support by the Spanish Ministerio Econom\'ia y Competitividad (MINECO) grant ESP2013-41268-R.
Part of this material is based upon work supported by the National
Science Foundation under Grant No. 1313484.
Part of this research was conducted by the Australian Research Council
Centre of Excellence for All-sky Astrophysics (CAASTRO), through project
number CE110001020.

This paper is based on observations collected with the 1.22m Galileo telescope of the Asiago Astrophysical Observatory, operated by the Department of Physics and Astronomy ``G. Galilei'' of the Universit\`a of Padova; the 1.82-m Copernico Telescope and the Schmidt 67/92cm of INAF-Asiago Observatory; the Italian TNG operated on the island of La Palma by the Fundaci\`on Galileo Galilei of the INAF (Istituto Nazionale di Astrofisica); the NOT, operated on the island of La Palma jointly by Denmark, Finland, Iceland, Norway and Sweden, in the Spanish Observatorio del Roque de los Muchachos of the Instituto de Astrof\`isica de Canarias; the Gran Telescopio Canarias (GTC), installed in the Spanish Observatorio del Roque de los Muchachos of the Instituto de Astrof\`isica de Canarias, in the Island of La Palma; the Liverpool Telescope is operated on the island of La Palma by Liverpool John Moores University in the Spanish Observatorio del Roque de los Muchachos of the Instituto de Astrofisica de Canarias with financial support from the UK Science and Technology Facilities Council; and the Telescopi Joan Or\`o of the Montsec Astronomical Observatory, which is owned by the Generalitat de Catalunya and operated by the Institute for Space Studies of Catalunya (IEEC). This paper is also based on observations made with the \textit{Swift} and LCOGT Observatories: we thank their respective staffs for excellent assistance. {\tt IRAF} is distributed by the National Optical Astronomy Observatory, which is operated by the Association of Universities for Research in Astronomy (AURA) under cooperative agreement with the National Science Foundation.

\newpage
\appendix
\subsection{Temperature from the ratio of \textnormal{[\ion{Ni}{II}] }lines}\label{sec: Ni/Ni}
In Section \ref{sec: Ni/Fe} we fixed the [\ion{Ni}{II}] $\lambda$7378 and $\lambda$7412 lines luminosity ratio following J15. However, as mentioned before, these two lines arise from two different atomic levels and their luminosity ratio is non-negligibly temperature dependent. This dependence, can be written as
\begin{equation}
\frac{L_{7412}}{L_{7378}}=\frac{n_8~A_{7412}~h\nu_{7412}~\beta_{7412}}{n_7~A_{7378}~h\nu_{7378}~\beta_{7378}}~~,
\end{equation}
where $n_7$ and $n_8$ are the number densities of level 7 and 8 (the levels from which the lines are from); $h$ is the Plank constant; $\nu$ is the frequency of the line; $\beta$ is the escape probability; $A_{7378}=0.23$~s$^{-1}$ and $A_{7412}=0.18$~s$^{-1}$ are atomic constants. The number densities ratio can be written as
\begin{equation}
\frac{n_8}{n_7}=\frac{\rm{e}^{-\Delta E_8/kT}g_8}{\rm{e}^{-\Delta E_7/kT}g_7}~~,
\end{equation}
where $\Delta E_7=13550$~cm$^{-1}$ and $\Delta E_8=14995$~cm$^{-1}$ are the level 7 and 8 energy respectively; $g_7=8$ and $g_8=6$ are the statistical weights of the levels. Assuming the lines to be optically thin (which they are according to the J15 model) then the escape probability are equal to 1 and thus we can write
\begin{equation}\label{eq: Ni/Ni}
\frac{L_{7412}}{L_{7378}}=0.584\rm{e}^{-0.18\rmn{eV}/kT}~~.
\end{equation}
Measuring the luminosity of the two Ni lines, then, we could have another estimate of the temperature. So we untied the bond on the two Ni lines luminosity ratios fixed in the previous analysis and redid the fit (note that doing this we added a parameter to the fit). We inferred a ratio of $L_{7412}/L_{7378}\simeq0.49$ (compare to the previous fixed 0.31) which gives the extremely high value of $T\simeq11600\pm8500$~K. This happens because Equation \ref{eq: Ni/Ni} has the shape of an hyperbola with an horizontal asymptote at $L_{7412}/L_{7378}=0.584$; then when this ratio is above 0.4 (which happens roughly at \mbox{$T=6000$~K}) the temperature starts to raise exponentially. Moreover for values \mbox{$L_{7412}/L_{7378}>0.584$} the temperatures become negative.

SN 2012ec had a very weak [\ion{Ca}{II}] which made the [\ion{Ni}{II}] $\lambda$7378 line the most prominent feature of this spectral region. In our case the [\ion{Ca}{II}] doublet was much more intense, and the whole structure was still too blended in order to easily detach every single feature, despite the low number of free parameters in the fit. This resulted in an unreliable measurement of the flux of the Ni lines and therefore in an unreliable estimate of the temperature $T$ with this method.


\section[Data]{Data}\label{Data}
Here we report the complete dataset of our measurements.

\begin{table*}
\begin{minipage}{140mm}
\renewcommand{\thefootnote}{\fnsymbol{footnote}}\caption{ugriz photometry}
\begin{tabular}{ccccccc}
\hline \\ 
MJD & u & g & r & i & z & Telescope\\ 
\hline \\ 
56673.47 & $-$ & 14.73 (0.04) & 14.79 (0.03) & 14.92 (0.03) & $-$ & LCOGT 1m\\ 
56674.31 & $-$ & 14.69 (0.01) & 14.76 (0.01) & 14.87 (0.03) & $-$ & LCOGT 1m\\ 
56675.53 & $-$ & 14.61 (0.02) & 14.63 (0.02) & 14.68 (0.01) & $-$ & LCOGT 1m\\ 
56677.25 & $-$ & 14.58 (0.02) & 14.52 (0.02) & 14.54 (0.03) & $-$ & LCOGT 1m\\ 
56678.29 & $-$ & 14.57 (0.02) & 14.49 (0.02) & 14.51 (0.03) & $-$ & LCOGT 1m\\ 
56679.30 & $-$ & 14.54 (0.02) & 14.41 (0.02) & 14.41 (0.03) & $-$ & LCOGT 1m\\ 
56682.21 & $-$ & 14.56 (0.03) & 14.36 (0.03) & 14.33 (0.03) & $-$ & LCOGT 1m\\ 
56683.27 & $-$ & 14.57 (0.04) & 14.34 (0.03) & 14.27 (0.03) & $-$ & LCOGT 1m\\ 
56684.26 & $-$ & 14.56 (0.04) & 14.34 (0.02) & 14.26 (0.04) & $-$ & LCOGT 1m\\ 
56685.19 & 15.82 (0.03) & 14.61 (0.03) & 14.39 (0.02) & 14.30 (0.02) & 14.28 (0.01) & LT\\ 
56689.52 & $-$ & 14.80 (0.03) & 14.44 (0.03) & 14.34 (0.02) & $-$ & LCOGT 1m\\ 
56693.21 & $-$ & 14.98 (0.03) & 14.48 (0.03) & 14.43 (0.04) & $-$ & LCOGT 1m\\ 
56695.15 & 16.90 (0.02) & 15.10 (0.03) & 14.59 (0.02) & 14.47 (0.03) & 14.41 (0.02) & LT\\ 
56697.03 & 17.05 (0.03) & 15.21 (0.03) & 14.59 (0.03) & 14.52 (0.02) & 14.40 (0.02) & LT\\ 
56697.38 & $-$ & 15.17 (0.04) & 14.62 (0.03) & 14.51 (0.03) & $-$ & LCOGT 1m\\ 
56699.00 & 17.26 (0.04) & 15.21 (0.04) & 14.65 (0.04) & 14.56 (0.03) & 14.48 (0.04) & LT\\ 
56701.14 & $-$ & 15.40 (0.03) & 14.72 (0.02) & 14.59 (0.03) & $-$ & LCOGT 1m\\ 
56701.97 & 17.70 (0.02) & 15.41 (0.03) & 14.73 (0.02) & 14.65 (0.03) & 14.61 (0.03) & LT\\ 
56702.22 & $-$ & 15.41 (0.05) & 14.76 (0.04) & 14.63 (0.04) & $-$ & LCOGT 1m\\ 
56704.21 & $-$ & 15.51 (0.06) & 14.84 (0.04) & 14.72 (0.03) & $-$ & LCOGT 1m\\ 
56709.14 & $-$ & 15.78 (0.03) & 14.94 (0.03) & 14.83 (0.04) & $-$ & LCOGT 1m\\ 
56717.17 & $-$ & 16.09 (0.06) & 15.17 (0.07) & 15.04 (0.05) & $-$ & LCOGT 1m\\ 
56721.47 & $-$ & 16.22 (0.02) & 15.24 (0.03) & 15.14 (0.03) & $-$ & LCOGT 1m\\ 
56733.19 & $-$ & 16.57 (0.03) & 15.46 (0.02) & 15.35 (0.03) & $-$ & LCOGT 1m\\ 
56743.14 & $-$ & 16.86 (0.06) & 15.58 (0.06) & 15.53 (0.05) & $-$ & LCOGT 1m\\ 
56745.11 & $-$ & 16.92 (0.04) & 15.63 (0.03) & 15.48 (0.03) & $-$ & LCOGT 1m\\ 
56749.22 & $-$ & 17.17 (0.05) & 15.83 (0.06) & 15.68 (0.04) & $-$ & LCOGT 1m\\ 
56751.39 & $-$ & 17.40 (0.02) & 15.97 (0.02) & 15.81 (0.02) & $-$ & LCOGT 1m\\ 
56757.21 & $-$ & 18.12 (0.05) & 16.60 (0.06) & 16.45 (0.04) & $-$ & LCOGT 1m\\ 
56759.32 & $-$ & 18.38 (0.05) & 16.76 (0.03) & 16.64 (0.03) & $-$ & LCOGT 1m\\ 
56762.29 & $-$ & 18.59 (0.07) & 17.03 (0.03) & 16.90 (0.03) & $-$ & LCOGT 1m\\ 
56764.34 & $-$ & 18.62 (0.05) & 17.00 (0.03) & 16.89 (0.03) & $-$ & LCOGT 1m\\ 
56785.22 & $-$ & 18.96 (0.24) & 17.37 (0.18) & 17.33 (0.25) & $-$ & LCOGT 1m\\ 
56791.24 & $-$ & 19.09 (0.21) & 17.38 (0.07) & 17.36 (0.12) & $-$ & LCOGT 1m\\ 
56795.24 & $-$ & 19.10 (0.19) & 17.54 (0.08) & 17.52 (0.10) & $-$ & LCOGT 1m\\ 
56805.22 & $-$ & 19.34 (0.17) & 17.68 (0.07) & 17.73 (0.11) & $-$ & LCOGT 1m\\ 
56820.25 & $-$ & 19.57 (0.15) & 17.75 (0.03) & 17.95 (0.03) & $-$ & LCOGT 2m\\ 
56822.32 & $-$ & 19.53 (0.51) & 17.84 (0.23) & 17.96 (0.22) & $-$ & LCOGT 2m\\ 
56835.29 & $-$ & 19.65 (0.27) & $-$ & $-$ & $-$ & LCOGT 2m\\ 
56837.27 & $-$ & 19.86 (0.12) & 18.08 (0.03) & 18.34 (0.04) & $-$ & LCOGT 2m\\ 
56838.28 & $-$ & 19.90 (0.13) & 18.06 (0.05) & 18.27 (0.06) & $-$ & LCOGT 2m\\ 
56840.28 & $-$ & 19.98 (0.30) & 18.09 (0.09) & 18.32 (0.11) & $-$ & LCOGT 2m\\ 
56843.26 & $-$ & 19.62 (0.52) & 18.13 (0.08) & 18.52 (0.18) & $-$ & LCOGT 2m\\ 
56959.06 & $-$ & 21.07 (0.19) & 20.20 (0.11) & 20.14 (0.17) & $-$ & 1.82m\\ 
57077.00 & $-$ & $<$ 22.6 & $<$ 22.4 & $<$ 22.1 & $-$ & TNG\\ 
\hline \\ 
\label{tab: ugriz}
\end{tabular}
\end{minipage}
\end{table*}

\begin{table*}
\begin{minipage}{140mm}
\caption{UBVRI photometry.}
\begin{tabular}{ccccccc}
\hline \\ 
MJD & U & B & V & R & I & Telescope\\ 
\hline \\ 
56663.23 & $-$ & $-$ & $-$ & $<$ 17.0 & $-$ & Wiggins\\ 
56671.32 & $-$ & $-$ & $-$ & 16.36 (0.19) & $-$ & Wiggins\\ 
56672.76 & 13.89 (0.05) & 15.06 (0.05) & 15.07 (0.05) & $-$ & $-$ & \textit{Swift}\\ 
56673.02 & 13.89 (0.04) & 15.03 (0.04) & 15.05 (0.04) & $-$ & $-$ & TNG\\ 
56673.31 & $-$ & $-$ & $-$ & 14.63 (0.27) & $-$ & Wiggins\\ 
56673.46 & $-$ & 14.90 (0.03) & 14.80 (0.02) & $-$ & $-$ & LCOGT 1m\\ 
56674.19 & 13.71 (0.03) & 14.85 (0.03) & 14.76 (0.06) & $-$ & $-$ & \textit{Swift}\\ 
56674.27 & $-$ & $-$ & $-$ & 14.61 (0.26) & $-$ & Wiggins\\ 
56674.29 & $-$ & 14.86 (0.02) & 14.74 (0.02) & $-$ & $-$ & LCOGT 1m\\ 
56675.25 & $-$ & $-$ & $-$ & 14.51 (0.11) & $-$ & Wiggins\\ 
56675.52 & $-$ & 14.77 (0.02) & 14.63 (0.03) & $-$ & $-$ & LCOGT 1m\\ 
56676.13 & 13.76 (0.01) & 14.84 (0.01) & 14.63 (0.01) & $-$ & $-$ & \textit{Swift}\\ 
56676.36 & $-$ & $-$ & $-$ & 14.39 (0.26) & $-$ & Wiggins\\ 
56677.24 & $-$ & 14.79 (0.02) & 14.50 (0.03) & $-$ & $-$ & LCOGT 1m\\ 
56677.35 & $-$ & $-$ & $-$ & 14.35 (0.24) & $-$ & Wiggins\\ 
56677.77 & 13.86 (0.03) & 14.87 (0.03) & 14.62 (0.04) & $-$ & $-$ & \textit{Swift}\\ 
56678.28 & $-$ & 14.77 (0.03) & 14.52 (0.02) & $-$ & $-$ & LCOGT 1m\\ 
56678.36 & $-$ & $-$ & $-$ & 14.36 (0.44) & $-$ & Wiggins\\ 
56678.91 & 13.91 (0.03) & 14.80 (0.03) & 14.56 (0.03) & $-$ & $-$ & \textit{Swift}\\ 
56678.97 & $-$ & 14.72 (0.22) & 14.44 (0.26) & 14.26 (0.31) & 13.98 (0.29) & Schmidt\\ 
56679.22 & $-$ & $-$ & $-$ & 14.28 (0.12) & $-$ & Wiggins\\ 
56679.29 & $-$ & 14.73 (0.03) & 14.47 (0.03) & $-$ & $-$ & LCOGT 1m\\ 
56680.03 & $-$ & 14.66 (0.32) & 14.43 (0.28) & 14.23 (0.46) & 14.02 (0.39) & Schmidt\\ 
56680.11 & 13.99 (0.02) & 14.78 (0.05) & 14.46 (0.06) & $-$ & $-$ & \textit{Swift}\\ 
56681.32 & $-$ & $-$ & $-$ & 14.29 (0.35) & $-$ & Wiggins\\ 
56681.57 & 14.05 (0.02) & 14.86 (0.02) & 14.54 (0.02) & $-$ & $-$ & \textit{Swift}\\ 
56682.11 & 14.07 (0.03) & 14.81 (0.03) & 14.41 (0.02) & 14.23 (0.03) & 14.02 (0.04) & TJO\\ 
56682.21 & $-$ & 14.78 (0.03) & 14.43 (0.03) & $-$ & $-$ & LCOGT 1m\\ 
56682.94 & 14.20 (0.11) & 14.99 (0.03) & 14.56 (0.03) & 14.24 (0.03) & 13.94 (0.05) & 1.82m\\ 
56683.26 & $-$ & 14.82 (0.03) & 14.41 (0.03) & $-$ & $-$ & LCOGT 1m\\ 
56684.26 & $-$ & 14.84 (0.04) & 14.46 (0.03) & $-$ & $-$ & LCOGT 1m\\ 
56685.19 & $-$ & 14.89 (0.03) & 14.43 (0.03) & $-$ & $-$ & LT\\ 
56686.38 & 14.52 (0.02) & 15.05 (0.02) & 14.60 (0.02) & $-$ & $-$ & \textit{Swift}\\ 
56688.03 & 14.79 (0.15) & 15.23 (0.37) & $-$ & 14.25 (0.19) & 13.92 (0.24) & TJO\\ 
56689.03 & 14.72 (0.03) & 15.14 (0.08) & 14.74 (0.07) & 14.29 (0.06) & 13.98 (0.12) & TJO\\ 
56689.51 & $-$ & 15.09 (0.03) & 14.59 (0.02) & $-$ & $-$ & LCOGT 1m\\ 
56690.09 & 14.82 (0.26) & 15.22 (0.03) & 14.58 (0.04) & 14.24 (0.04) & 13.98 (0.06) & TJO\\ 
56691.05 & 15.00 (0.04) & 15.41 (0.07) & 14.73 (0.06) & 14.30 (0.05) & 14.01 (0.12) & TJO\\ 
56693.20 & $-$ & 15.30 (0.02) & 14.72 (0.02) & $-$ & $-$ & LCOGT 1m\\ 
56695.15 & $-$ & 15.46 (0.03) & 14.80 (0.02) & $-$ & $-$ & LT\\ 
56695.85 & 15.63 (0.02) & 15.60 (0.01) & 14.96 (0.02) & $-$ & $-$ & \textit{Swift}\\ 
56697.03 & $-$ & 15.53 (0.03) & 14.81 (0.03) & $-$ & $-$ & LT\\ 
56697.08 & 15.53 (0.62) & 15.77 (0.04) & 14.89 (0.03) & 14.44 (0.03) & 14.11 (0.10) & TJO\\ 
56697.36 & $-$ & 15.52 (0.04) & 14.85 (0.02) & $-$ & $-$ & LCOGT 1m\\ 
56699.00 & $-$ & 15.69 (0.03) & 14.88 (0.03) & $-$ & $-$ & LT\\ 
56699.21 & $-$ & 15.75 (0.04) & 14.93 (0.04) & 14.53 (0.05) & 14.15 (0.07) & TJO\\ 
56700.07 & $-$ & 15.85 (0.05) & 15.04 (0.05) & 14.61 (0.05) & 14.21 (0.06) & TJO\\ 
56701.03 & $-$ & 15.77 (0.10) & 14.93 (0.08) & 14.66 (0.02) & 14.05 (0.13) & NOT (StanCam)\\ 
56701.14 & $-$ & 15.81 (0.04) & 14.99 (0.02) & $-$ & $-$ & LCOGT 1m\\ 
56701.97 & $-$ & 15.88 (0.03) & 15.00 (0.03) & $-$ & $-$ & LT\\ 
56702.22 & $-$ & 15.85 (0.04) & 15.02 (0.03) & $-$ & $-$ & LCOGT 1m\\ 
56704.20 & $-$ & 15.97 (0.05) & 15.12 (0.04) & $-$ & $-$ & LCOGT 1m\\ 
56709.13 & $-$ & 16.23 (0.03) & 15.26 (0.02) & $-$ & $-$ & LCOGT 1m\\ 
56715.25 & 17.39 (0.02) & 16.67 (0.02) & 15.46 (0.02) & 14.90 (0.02) & 14.49 (0.02) & TNG\\ 
56717.16 & $-$ & 16.60 (0.06) & 15.56 (0.07) & $-$ & $-$ & LCOGT 1m\\ 
56721.46 & $-$ & 16.76 (0.03) & 15.61 (0.02) & $-$ & $-$ & LCOGT 1m\\ 
56727.11 & $-$ & 17.15 (0.06) & 15.81 (0.05) & 15.09 (0.04) & 14.78 (0.04) & 1.82m\\ 
56727.89 & $-$ & 17.08 (0.05) & 15.68 (0.12) & 15.12 (0.17) & 14.84 (0.14) & Schmidt\\ 
56730.97 & $-$ & 17.31 (0.04) & 15.82 (0.02) & 15.13 (0.02) & 14.84 (0.02) & NOT (StanCam)\\ 
56733.43 & $-$ & 17.22 (0.04) & 15.95 (0.03) & $-$ & $-$ & LCOGT 1m\\ 
56733.88 & $-$ & 17.25 (0.11) & 15.73 (0.14) & 15.13 (0.19) & 14.79 (0.12) & Schmidt\\ 
\hline \\ 
\label{tab: UBVRI}
\end{tabular}
\end{minipage}
\end{table*}

\begin{table*}
\begin{minipage}{140mm}
\renewcommand{\thefootnote}{\fnsymbol{footnote}}\contcaption{UBVRI photometry.}
\begin{tabular}{ccccccc}
\hline \\ 
MJD & U & B & V & R & I & Telescope\\ 
\hline \\ 

56737.01 & $-$ & 17.39 (0.05) & 15.95 (0.10) & 15.20 (0.25) & 14.93 (0.21) & Schmidt\\ 
56743.13 & $-$ & 17.53 (0.06) & 16.14 (0.05) & $-$ & $-$ & LCOGT 1m\\ 
56746.26 & $-$ & 17.70 (0.04) & 16.32 (0.04) & $-$ & $-$ & LCOGT 1m\\ 
56747.13 & 18.97 (0.23) & 17.90 (0.07) & 16.26 (0.09) & 15.42 (0.03) & 15.26 (0.12) & 1.82m\\ 
56749.38 & $-$ & 17.93 (0.05) & 16.42 (0.03) & $-$ & $-$ & LCOGT 1m\\ 
56754.96 & $-$ & 18.52 (0.08) & 16.92 (0.05) & 15.93 (0.03) & 15.52 (0.06) & 1.82m\\ 
56755.14 & $-$ & 18.49 (0.06) & 17.02 (0.02) & $-$ & $-$ & LCOGT 1m\\ 
56757.24 & $-$ & 18.73 (0.08) & 17.31 (0.06) & $-$ & $-$ & LCOGT 1m\\ 
56759.35 & $-$ & 18.97 (0.13) & 17.58 (0.07) & $-$ & $-$ & LCOGT 1m\\ 
56762.28 & $-$ & $<$ 18.9 & 17.92 (0.12) & $-$ & $-$ & LCOGT 1m\\ 
56764.04 & $-$ & 19.22 (0.14) & 17.91 (0.06) & 17.00 (0.03) & 16.32 (0.05) & NOT (StanCam)\\ 
56764.32 & $-$ & 19.43 (0.11) & 18.04 (0.04) & $-$ & $-$ & LCOGT 1m\\ 
56764.88 & $-$ & 19.25 (0.09) & 17.98 (0.12) & 16.94 (0.20) & 16.27 (0.14) & Schmidt\\ 
56785.21 & $-$ & $<$ 18.6 & 18.41 (0.26) & $-$ & $-$ & LCOGT 1m\\ 
56791.22 & $-$ & 19.72 (0.22) & 18.46 (0.09) & $-$ & $-$ & LCOGT 1m\\ 
56795.22 & $-$ & 19.87 (0.14) & 18.52 (0.07) & $-$ & $-$ & LCOGT 1m\\ 
56797.00 & $-$ & 19.69 (0.09) & 18.53 (0.06) & 17.48 (0.01) & 16.89 (0.05) & NOT (StanCam)\\ 
56805.19 & $-$ & $<$ 19.4 & 18.80 (0.19) & $-$ & $-$ & LCOGT 1m\\ 
56820.18 & $-$ & $<$ 19.1 & 18.79 (0.23) & $-$ & $-$ & LCOGT 1m\\ 
56959.04 & $-$ & $<$ 20.9 & 21.29 (0.27) & $-$ & $-$ & 1.82m\\ 
57047.04 & $-$ & $-$ & $<$ 21.9 & $<$ 20.7 & $<$ 20.5 & NOT (ALFOSC)\\ 
\hline \\ 
\label{tab: UBVRI}
\end{tabular}
\end{minipage}
\end{table*}

\begin{table*}
\begin{minipage}{140mm}
\renewcommand{\thefootnote}{\fnsymbol{footnote}}\caption{NIR photometry.}
\begin{tabular}{ccccc}
\hline \\ 
MJD & J & H & K & Telescope\\ 
\hline \\ 
56701.00 & 13.98 (0.34) & 13.62 (0.37) & 13.54 (0.36) & NOT (NOTCam)\\ 
56731.00 & 14.08 (0.25) & 14.28 (0.14) & 13.67 (0.29) & NOT (NOTCam)\\ 
56764.06 & 15.10 (0.25) & 15.23 (0.32) & 14.77 (0.39) & NOT (NOTCam)\\ 
56797.03 & 16.24 (0.23) & 15.64 (0.36) & 14.93 (0.36) & NOT (NOTCam)\\ 
56847.89 & 17.45 (0.31) & 16.60 (0.31) & 15.39 (0.28) & NOT (NOTCam)\\ 
\hline \\ 
\label{tab: JHK}
\end{tabular}
\end{minipage}
\end{table*}

\begin{table*}
\begin{minipage}{140mm}
\renewcommand{\thefootnote}{\fnsymbol{footnote}}\caption{UV photometry.}
\begin{tabular}{ccccc}
\hline \\ 
MJD & UVW2 & UVM2 & UVW1 & Telescope\\ 
\hline \\ 
56672.95 & 14.15 (0.03) & 14.09 (0.05) & 13.94 (0.02) & \textit{Swift}\\ 
56674.12 & 14.25 (0.02) & 14.21 (0.05) & 13.94 (0.01) & \textit{Swift}\\ 
56676.13 & 14.75 (0.01) & 14.45 (0.01) & 14.11 (0.01) & \textit{Swift}\\ 
56677.87 & 15.14 (0.04) & 14.91 (0.03) & 14.36 (0.03) & \textit{Swift}\\ 
56679.00 & 15.36 (0.04) & 15.15 (0.03) & 14.49 (0.03) & \textit{Swift}\\ 
56680.11 & 15.66 (0.03) & $-$ & 14.74 (0.06) & \textit{Swift}\\ 
56681.54 & 15.82 (0.02) & 15.64 (0.02) & 14.87 (0.02) & \textit{Swift}\\ 
56686.38 & 16.80 (0.03) & 16.64 (0.03) & 15.63 (0.02) & \textit{Swift}\\ 
56695.85 & 18.42 (0.05) & 18.68 (0.06) & 17.10 (0.02) & \textit{Swift}\\ 
\hline \\ 
\label{tab: ADS}
\end{tabular}
\end{minipage}
\end{table*}

\begin{landscape}
\begin{table}
\caption{Optical spectroscopy data.}
\begin{tabular}{cccccccccccc}
\hline \\ 
Epoch (MJD) & Telescope & Temp. & H$\alpha$ (emi.) & H$\alpha$ (abs.) & H$\beta$ & H$\gamma$ & He I+Na I $\lambda5876$ & Fe II $\lambda5018$ & Fe II $\lambda5169$ & Sc II $\lambda5527$ & Sc II $\lambda6245$ \\
 & & [K] & [km s$^{-1}$]& [km s$^{-1}$] & [km s$^{-1}$] & [km s$^{-1}$] & [km s$^{-1}$] & [km s$^{-1}$] & [km s$^{-1}$] & [km s$^{-1}$] & [km s$^{-1}$] \\
\hline \\ 
2014-01-14 (56672.13) & 1.22m & 23069 & $-$ & $-$ & $-$ & $-$ & $-$ & $-$ & $-$ & $-$ & $-$ \\
2014-01-15 (56672.91) & 1.22m & 15835 & $-$ & $-$ & $-$ & $-$ & $-$ & $-$ & $-$ & $-$ & $-$ \\
2014-01-21 (56678.92) & 1.22m & 12151 & $-$ & $-$ & $-$ & $-$ & $-$ & $-$ & $-$ & $-$ & $-$ \\
2014-01-25 (56682.92) & 1.82m & 12192 & $-$ & $-$ & $-$ & $-$ & 8421 (81) & 7914 (74) & $-$ & $-$ & $-$ \\
2014-01-28 (56686.21) & 1.22m & 9849 & $-$ & 8262 (85) & 9542 (121) & 10003 (397) & 8325 (97) & 7906 (152) & 8239 (181) & $-$ & $-$ \\
2014-02-06 (56694.94) & 1.22m & 7749 & $-$ & 8598 (30) & 9382 (32) & 9487 (67) & 7823 (40) & $-$ & 7651 (45) & 8740 (58) & 8326 (59) \\
2014-02-17 (56706.03) & 1.22m & 7158 & 8577 & 8311 (42) & 8378 (51) & 7948 (178) & 7282 (81) & 7339 (55) & 6503 (78) & 7460 (98) & 7101 (260) \\
2014-02-19 (56708.02) & TNG & 6587 & 8754 & 8073 (10) & 8281 (113) & 7978 (96) & 7164 (23) & 7125 (33) & 6574 (45) & 7468 (34) & 7118 (37) \\
2014-02-26 (56715.19) & TNG & 5891 & 8422 & 7849 (13) & 7739 (106) & 7260 (68) & 6758 (22) & 6847 (80) & 6107 (60) & 6989 (50) & 6638 (79) \\
2014-03-06 (56722.89) & 1.22m & 6918 & 8642 & 7738 (24) & 7241 (28) & 6428 (202) & 6344 (27) & 6239 (287) & 5640 (82) & 6791 (77) & 6381 (94) \\
2014-03-10 (56727.08) & 1.82m & 6595 & 8409 & 7427 (13) & 6846 (104) & $-$ & 6093 (204) & 5103 (374) & 5153 (237) & $-$ & $-$ \\
2014-03-17 (56733.80) & 1.22m & 5970 & 8640 & 7426 (35) & 6842 (37) & 6558 (170) & 5684 (46) & 4795 (197) & 4711 (86) & $-$ & $-$ \\
2014-03-30 (56747.06) & 1.82m & 5983 & 8363 & 7065 (12) & 6132 (76) & 5106 (194) & 4632 (36) & 3007 (44) & 3733 (318) & $-$ & 2729 (653) \\
2014-04-07 (56754.99) & 1.82m & 5823 & 8029 & 6973 (46) & 5774 (93) & 4730 (224) & 4348 (25) & 2685 (43) & 3427 (135) & $-$ & 2218 (147) \\
2014-04-25 (56772.93) & NOT & 5399 & 6229 & $-$ & 5573 (70) & 4376 (247) & 4180 (25) & 3156 (288) & 3280 (100) & $-$ & $-$ \\
2014-05-11 (56788.91) & TNG & 5422 & 5530 & $-$ & 5347 (75) & 3697 (207) & 3865 (29) & 2780 (130) & 3288 (63) & $-$ & $-$ \\
2014-05-29 (56806.90) & TNG & 5272 & 4979 & $-$ & 5169 (81) & 3540 (288) & 3666 (43) & 2991 (246) & 3328 (81) & $-$ & $-$ \\
2014-07-18 (56856.89) & TNG & $-$ & 4306 & $-$ & $-$ & $-$ & 3443 (53) & $-$ & $-$ & $-$ & $-$ \\
2014-12-20 (57012.09) & GTC & $-$ & 4647 & $-$ & $-$ & $-$ & $-$ & $-$ & $-$ & $-$ \\
\hline \\ 
\label{tab: spec}
\end{tabular}
\end{table}
\end{landscape}

\bsp

\label{lastpage}

\end{document}